\documentclass[journal]{IEEEtran}

\usepackage{amsmath,amsfonts}
\usepackage{algorithmic}
\usepackage{algorithm}
\usepackage{array}
\usepackage[caption=false]{subfig}
\usepackage{textcomp}
\usepackage{stfloats}
\usepackage{url}
\usepackage{verbatim}
\usepackage{graphicx}
\usepackage{cite,citesort}
\usepackage{color}
\usepackage{amssymb}
\usepackage{enumerate}
\usepackage{multirow}
\usepackage{multicol}

\newtheorem{theorem}{Theorem}

\newtheorem{lemma}{Lemma}

\newtheorem{remark}{Remark}

\DeclareMathOperator{\sinc}{sinc}

\begin{document}

\title{{Spectrum and Orthogonality of} Orthogonal Delay-Doppler Division Multiplexing Modulation Waveforms}

\author{
Akram~Shafie, \IEEEmembership{Member, IEEE,}~Jun Tong, \IEEEmembership{Senior Member, IEEE,}~Jinhong~Yuan, \IEEEmembership{Fellow, IEEE,} Taka Sakurai, \IEEEmembership{Senior Member, IEEE,} Paul Fitzpatrick, \IEEEmembership{Senior Member, IEEE,} Yuting Fang, \IEEEmembership{Member, IEEE,} and Yixuan Xie. \IEEEmembership{Member, IEEE}

\thanks{A. Shafie, J. Yuan, and Y. Xie are with the School of Electrical Engineering and Telecommunications,  University of New South Wales, Sydney, NSW, 2052, Australia (e-mail: \{akram.shafie, j.yuan, yixuan.xie\}@unsw.edu.au).} \thanks{J. Tong is with the School of Electrical, Computer and Telecommunications Engineering, University of Wollongong, Wollongong, NSW 2522, Australia (e-mail: jtong@uow.edu.au).}
\thanks{ T. Sakurai, P. Fitzpatrick, and Y. Fang are with Telstra Limited, Melbourne, Australia (e-mail: \{paul.g.fitzpatrick,~yuting.fang\}@team.telstra.com).}
\thanks{
A preliminary version of this work was presented at the 2025 IEEE  ICC~\cite{AS_2025_C_ICC_ODDMSpectrum}.}
}

\markboth{Submitted to IEEE Transactions on Wireless Communications}%
{Shafie \MakeLowercase{\textit{et al.}}: On the Spectral Response, Orthogonality, and Implementation of Orthogonal Delay-Doppler Division Multiplexing Modulation Waveforms}

\maketitle

\begin{abstract}
Orthogonal delay-Doppler (DD) division multiplexing (ODDM) modulation has recently emerged as a promising paradigm for ensuring reliable communications in doubly-selective channels. This work investigates the spectra and orthogonality characteristics of analog (direct) and {{approximate digital}} implementations of ODDM systems. We first determine the time and frequency domain representations of the basis functions for waveform in analog and {{approximate digital}} ODDM systems. Thereafter, we derive their power spectral densities and show that while the spectrum of analog ODDM waveforms exhibits a step-wise behavior in its transition regions, the spectrum of {{approximate digital}} ODDM waveforms is confined to that of the ODDM sub-pulse. Next, we prove the orthogonality characteristics of {{approximate digital}} ODDM waveforms and show that, unlike analog ODDM waveforms, the {{approximate digital}} ODDM waveforms satisfy orthogonality without the need of additional time domain resources. Additionally, we examine the similarities and differences that implementations of {{approximate digital}} ODDM share with the other variants of DD modulations, focusing on the domain changes the symbols undergo, the type of pulse shaping and windowing used, and the domains and the sequence in which they are performed. Finally, we present numerical results to %
validate our findings and 
draw further insights.
\end{abstract}


\section{Introduction}

\IEEEPARstart{M}{ulticarrier (MC)} modulations have been at the forefront of wireless systems for the last couple of decades~\cite{1997_Hass_TFLocalization,2009_Weinstein_HistoryofOFDM}. As an MC modulation scheme on the time-frequency (TF) plane, orthogonal frequency division multiplexing (OFDM) modulation has dominated 4G, 5G, and Wi-Fi, thanks to its simple transceiver implementation and the utilization of {sub-channel} single-tap equalization at the receiver for linear time-invariant (LTI)  channels~\cite{2018_ComStandardMag_OFDMNumerology}.


Despite its advantages, it remains uncertain whether OFDM can support reliable communication in some of the envisioned 6G environments, such as high-mobility, underwater, and high-frequency communication scenarios, and integrated sensing and communication scenarios~\cite{2021_WCM_JH_OTFS,2017_WCNC_OTFS_Haddani,AS_2024_TCOM_OTFSOFDMCoexistence}. These {environments} present several challenges, including severe Doppler spread, which compromises the orthogonality characteristics of OFDM~\cite{CAS_2025_J_WCL_Yiyan_CSFInspiredCTFEstimationforOFDM}. To address these issues in linear time-variant (LTV) channels, a promising alternative to OFDM has recently emerged in the form of another MC modulation scheme, termed orthogonal delay-Doppler (DD) division multiplexing (ODDM) modulation~\cite{2022_ICC_JH_ODDM,2022_TWC_JH_ODDM}.


As an MC modulation on the DD plane, the ODDM maps the information-bearing symbols into the DD plane. {Then, it utilizes the DD orthogonal pulse (DDOP) to obtain the transmit ODDM {waveform}. The DDOP serves as the transmit pulse of MC modulation on the DD plane and can be constructed by concatenating multiple square root Nyquist (SRN) sub-pulses. Thanks to the orthogonality characteristics of DDOP with respect to (w.r.t.) the fine resolutions of the DD plane,} ODDM effectively couples the modulated {waveform} with the DD domain representation of the 
LTV channels, which is known for its inherent stability and sparsity properties~\cite{2022_TWC_JH_ODDM}. This coupling minimizes interference and establishes a compact input-output (I/O) relation, enabling accurate channel estimation with minimal pilot overhead, 
and simplified equalization  
over LTV channels~\cite{2024_TCOM_JunTong_ODDMoverPhysicalChannels}.  


Two distinct {implementation} approaches for generating ODDM waveforms have been proposed~\cite{2022_ICC_JH_ODDM,2022_TWC_JH_ODDM}. The first{, which is the direct approach,} exploits the fact that ODDM is an MC modulation on the DD (fine-time and fine-frequency) plane. 
Consequently, 
the ODDM {waveform} is generated 
{using the MC modulator} as the sum of multiple fine-time (delay) and fine-frequency (Doppler) shifted versions of modulated DDOP (see (21) in\cite{2022_TWC_JH_ODDM}). 
{Given that DDOP has a well-defined time representation and spectrum~\cite{2023_ODDM_TCOM}, this approach aims to generate the ODDM waveform that adheres to the time and frequency resource constraints of communication channels.}
As this {implementation} approach requires multiple analog devices in its transceivers {(e.g., mixers and time division multiplexer (TMX))}, in this paper,  we refer to the corresponding system as \textit{analog {(direct)} ODDM system}, and {the resulting generated waveforms as \textit{analog ODDM waveforms}}~\cite{2022_TWC_JH_ODDM,2023_ODDM_TCOM}.



The connection between MC modulation/demodulation and the inverse discrete Fourier transform (IDFT)/discrete Fourier transform
(DFT) were discovered in \cite{1971_Weinstein_OFDM_usingDFT,1967_IEEETCT_zimmerman_OFDM_usingDFT}.   {Also, the  fast Fourier transform (FFT) and inverse  FFT (IFFT) algorithms were introduced in~\cite{1969_ACM_IFFT} to realize the DFT and IDFT.} These lead to the IFFT/FFT-based direct \emph{digital} implementation of MC modulation, which paved the way for simpler transceiver implementations in OFDM systems~\cite{2009_Weinstein_HistoryofOFDM}. Leveraging this connection for ODDM, {and introducing an approximation to eliminate the need for DDOP-based truncation/windowing  and time multiplexing, \cite{2022_ICC_JH_ODDM,2022_TWC_JH_ODDM} proposed a low complexity {approximate digital} implementation of the ODDM system.} 
The idea is to transform the information-bearing DD domain symbols into a time domain digital sequence and then to perform sample-wise pulse shaping based on the ODDM sub-pulse (see (18) in\cite{2022_TWC_JH_ODDM}). 
{Recognizing that most operations (e.g., symbol domain transformation and pulse shaping) in this implementation} approach function primarily in the digital domain, {and recognizing the involved approximation,} 
in this paper, 
we refer to the corresponding system as \textit{{approximate} digital ODDM system}, and the {resulting generated waveforms as \textit{{approximate} digital ODDM waveforms}}~\cite{2022_TWC_JH_ODDM,2023_ODDM_TCOM}.


Several studies have explored the characteristics of {analog and digital ODDM waveforms and their applicability} in different contexts. 
In particular,~\cite{2022_ICC_JH_ODDM,2022_TWC_JH_ODDM} proved the orthogonality characteristics of analog ODDM waveforms w.r.t. the delay (fine time) and Doppler (fine frequency) resolutions of ODDM. Moreover,~\cite{2022_GC_JH_ODDMPulse,2023_ODDM_TCOM} proposed a 
generalized analog ODDM {waveform} in which the sub-pulse duration was increased while introducing {additional time domain cyclic extensions.} 
Then,~\cite{2022_GC_JH_ODDMPulse,2023_ODDM_TCOM} proved the orthogonality characteristics of the generalized analog ODDM {waveform}.
Furthermore,~\cite{AS_2025_J_TCOM_ODDMTFLoc,2023_ODDM_TCOM} investigated the TF localization characteristics, the {Fourier representation/frequency domain representation}, and the bandwidth efficiency of the DDOP. 
Additionally,~\cite{2024_TCOM_JunTong_ODDMoverPhysicalChannels,CAS_2025_J_TWC_Jun_ChannelforODDMwithHighSpread} derived the I/O relation of {approximate digital} ODDM systems over general physical channels. 
Despite these preliminary {results}, various aspects of analog and {approximate digital} ODDM systems still remain underexplored and require further investigation.



First, although ODDM is an MC modulation in the DD plane, the academic and industry communities are more accustomed to understanding and altering the time and frequency resources of waveforms and signals. Therefore, to successfully comprehend ODDM and then implement it in future communication systems, it is crucial to understand the time and frequency domain occupancies of ODDM waveforms. Particularly,  the time and frequency domain representations of basis functions, the {spectra}, and  OOBE performances of analog and {approximate digital} ODDM waveforms need to be analytically examined. 
{This examination can pave the way for evaluating the compliance of ODDM waveforms with spectral masks defined by various standards, including those for future 6G systems~\cite{3GPPWaveformCandidates}.}

While the {frequency domain representation} of DDOP was derived in~\cite{2023_ODDM_TCOM}, this derivation was not extended to understand the power {spectra} of ODDM waveforms. Recently,~\cite{2023_Arman_PulseShapingforDDM,2023_ZakOTFS_Implementation_Macquarie} reported a step-wise power spectral density behavior for ODDM waveforms through numerical results. However, the analytical reason for this behavior has not been elucidated, nor has it been established whether this behavior is a characteristic of analog ODDM {waveform}, {approximate digital} ODDM waveforms, or both.


Second, while the orthogonality characteristics of analog ODDM waveforms have been studied in~\cite{2022_ICC_JH_ODDM,2022_TWC_JH_ODDM,2022_GC_JH_ODDMPulse,2023_ODDM_TCOM}, a similar investigation into the orthogonality characteristics of {approximate digital} ODDM waveforms is lacking. 
It has to be noted that due to differences in analog and {approximate digital} ODDM systems, there may be variations in their transmit signals, the symbols obtained after receiver processing, and orthogonality characteristics, necessitating separate investigations. 




{
Third, DD modulation—where information-bearing symbols are embedded in the DD domain and allowed to interact with the DD domain representation of doubly-selective LTV channels—was first introduced in the form of OTFS~\cite{2017_WCNC_OTFS_Haddani}. Later, the $T$-length rectangular pulse/window was proposed in \cite{2018_TWC_Viterbo_OTFS_InterferenceCancellation}  as the transmit pulse for this original OTFS proposal.} Inspired by this original OTFS proposal with rectangular pulse/window and ODDM, several other variants of DD modulation have been recently proposed. To the best of the authors' knowledge, these include
{discrete Zak transform-based OTFS with rectangular transmitter and receiver window \cite{2022_EmanuelBook_DDCom},} Zak-OTFS~\cite{2023_OTFS20_Paper1},  Zak-OTFS with Type 1 and 2 filters~\cite{2023_ZakOTFS_Implementation_Macquarie}, circular or linear pulse-shaped OTFS \cite{2023_Arman_PulseShapingforDDM}, {DD-OFDM~\cite{2023_OFDMbasedODDM_Macquarie_Final},} and OTFS via discrete-Zak transform and time and frequency windowing~\cite{2023_Shangyuan_ZakbasedPulseShapingforOTFS}. 
{While ODDM and these DD modulation variants share more similarities than differences, ODDM and these variants have typically been developed independently. As a result, cross-comparisons among ODDM and these variants are missing, meaning their relative strengths and weaknesses have not been thoroughly examined.}
Although \cite{2022_ICC_JH_ODDM,2023_ZakOTFS_Implementation_Macquarie,2023_Arman_PulseShapingforDDM} have numerically compared the {spectra} of ODDM with some of these variants and the original OTFS, a comprehensive investigation into the similarities and differences between these variants is still lacking. 

\color{black}

Moreover, as an alternative to ODDM and DD modulation, chirp-based MC modulation schemes—such as affine frequency division multiplexing (AFDM)~\cite{2023_TWC_AliBemani_AFDM} and orthogonal chirp division multiplexing (OCDM)~\cite{2016_TCOM_OCDM}—have recently been introduced.
In particular, AFDM enables effective channel path separation in the affine/discrete affine Fourier transform (DAFT) domain, allowing it to exploit full diversity and achieve reliable communication over LTV channels. It is worth noting that AFDM typically suffers from higher PAPR than ODDM/OTFS, and owing to its recent introduction, the body of research on AFDM is comparatively less developed in areas such as equalisation and channel estimation algorithms, as well as multiuser and MIMO system design, making it less ready for practical deployment. Nevertheless, AFDM offers key advantages compared to ODDM/OTFS such as reduced pilot overhead for channel estimation and strong backward compatibility with OFDM, which is especially important given the recent decision by standardisation bodies to continue with OFDM for 6G systems~\cite{2025_Savaux_AFDMvsOTFS,2025_Mag_IEEENetworks_AFDM,2025_Savaux_TCOM_AllDFT,2025_ChirpPermutedAFDM}.
Considering these, it is valuable to explore whether chirp-based designs share implementation similarities or exhibit fundamental differences compared to ODDM and DD modulation.

\color{black}


Recognizing the above gaps in the literature, in this work, we investigate analog and {approximate digital} ODDM systems, in terms of their implementation, time and frequency domain representations, as well as orthogonality characteristics of their waveforms, and similarities and differences that they share with other DD modulation variants. 
The main contributions of this work are summarized as follows:
\begin{itemize}
  \item We explicitly determine the time and frequency domain representations of the basis functions of analog and {approximate digital} ODDM waveforms. Based on this, we demonstrate that \color{black}in the time domain, the basis functions of both waveforms are characterized by a \textit{sub-pulse train} and a \textit{phase term}. While the sub-pulse train is identical for both, their phase terms differ. In the frequency domain, the basis functions of both waveforms can be described by an \textit{infinite train of sinc tones} and an \textit{envelope}. Although the infinite train of sinc tones is the same in both, their envelopes differ. \color{black}
  \item We derive the expected power spectral densities of the transmit analog and {approximate digital} ODDM waveforms. Based on this, we analytically show that the spectrum of analog ODDM waveforms exhibits a step-wise behavior in its transition regions, 
       whereas those of {approximate digital} ODDM waveforms are confined strictly to the spectrum of the sub-pulse. 
  \item We prove the orthogonality characteristics of the {approximate digital} ODDM waveforms. We show that, unlike analog ODDM waveforms, the {approximate digital} ODDM  waveforms can attain orthogonality w.r.t. delay and Doppler resolutions without necessitating {additional time domain cyclic extensions,} 
      even when the sub-pulse duration is high.
  
  \item We compare the implementation of ODDM with those of the other variants of DD modulation. We show that the similarities and differences between {{approximate digital} ODDM systems} and those of other DD modulation variants stem from the domain changes the symbols undergo, the type of pulse shaping and windowing used, and the domains and the sequence in which they are performed.
  \item Using numerical results, we 
  show that 
    {approximate digital} ODDM systems achieve a similar bit error rate compared to their analog counterpart while requiring significantly lower implementation complexity. 
\end{itemize}



{\textbf{Notations:}
We denote $\mathcal{S}_{\Psi_1,\Psi_2}=\{-\Psi_1, -\Psi_1+1,\cdots,\Psi_2-1\}$ as the index set, where 
$\Psi_1$ and $\Psi_2$ are non-negative integers. For example, $\mathcal{S}_{0,N}=\{0,1,\cdots,N-1\}$ and $\mathcal{S}_{\frac{N}{2},\frac{N}{2}}=\{-\frac{N}{2},-\frac{N}{2}+1,\cdots,0,\cdots,\frac{N}{2}-1\}$.
}
{Moreover, $[.]_N$ denotes the modulo-$N$ operation yielding results in the set $\mathcal{S}_{0,N}$. 
Furthermore, $\psi_{N}(n)$ denotes the reverse of the mod $N$ operation that yields results only in the set $\mathcal{S}_{\frac{N}{2},\frac{N}{2}}$.
 Thus,  $\psi_{N}(n)=n$ if $0\leqslant n \leqslant\frac{N}{2}-1$ and $\psi_{N}(n)=n-N$ if $\frac{N}{2}\leqslant n\leqslant N-1$. For example, $\psi_{N}(N/4)=N/4$ and $\psi_{N}(3N/4)=-N/4$.
 }

\section{Analog and {Approximate} Digital ODDM {Systems}}
\label{Sec:ODDMImplementation}


\subsection{Analog ODDM System}

{The analog (direct) ODDM system utilizes the MC modulator and demodulator in its transmitter and receiver, respectively~\cite{1997_Hass_TFLocalization}. }
The schematic representation of the analog ODDM system is shown in Fig.~\ref{Fig:A-ODDM}.

\subsubsection{Transmitter}



We consider the transmission of $MN$ information-bearing symbols within the allocated bandwidth $\frac{M}{T}$ and the transmission time $NT$. 
In ODDM, information-bearing symbols are mapped into the DD domain to obtain  $X_{\textrm{dD}}[m,n]$,
where $m$ and $n$ denote the delay and Doppler indices, respectively, with $m\in\mathcal{S}_{0,M}$ 
and $n\in\mathcal{S}_{0,N}$, 
and $M$ and $N$ denote the total number of delay and Doppler elements, respectively. 

\begin{figure*}[t]
\centering
\includegraphics[width=2\columnwidth]{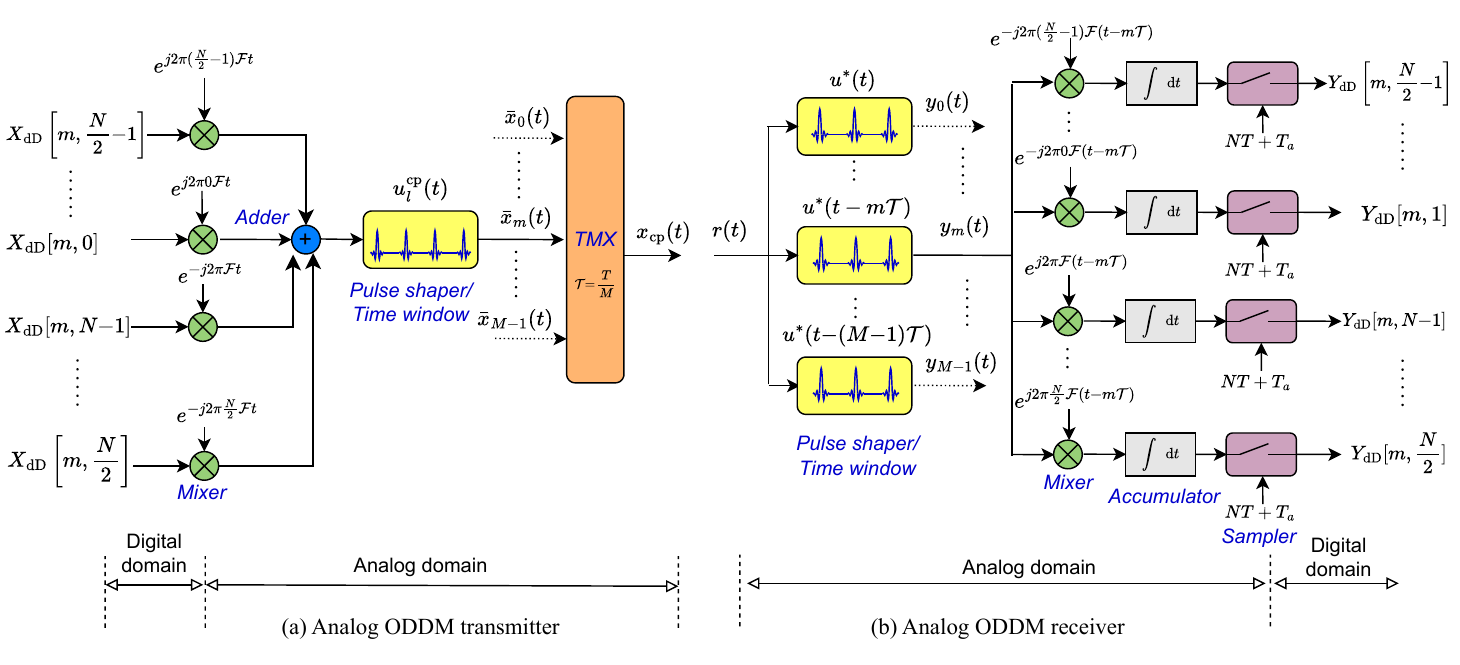}
\caption{Illustrations of the transmitter and the receiver of analog ODDM system.}\label{Fig:A-ODDM}
\end{figure*}


{As shown} in Fig.~\ref{Fig:A-ODDM}(a), the symbols $X_{\textrm{dD}}[m,n]$ are first modulated to the carriers with frequency separation of $\mathcal{F}$ using mixers/modulators, where 
$\mathcal{F}\triangleq\frac{1}{NT}$ denotes the Doppler (fine frequency) resolutions of ODDM, which is also the sub-carrier spacing of MC modulation \cite{2022_ICC_JH_ODDM,2022_TWC_JH_ODDM}. 
Thereafter, the resulting signals are aggregated and then truncated based on $u^{\textrm{cp}}_m(t)$ to obtain the $m$th ODDM symbol 
\begin{align} \label{Equ:xlt_A-ODDM}
\bar{x}_{m}(t)\!=\!u^{\textrm{cp}}_m(t)\!\!\!\sum_{n=-\frac{N}{2}}^{\frac{N}{2}-1} \!\!\!X_{\mathrm{dD}}\left[m,\left[n\right]_N\right]e^{j2\pi n \mathcal{F}t},~\forall m\in\mathcal{S}_{0,M},\!\!
\end{align}
where $u^{\textrm{cp}}_m(t)$ denotes the CP-appended version of the fundamental/prototype pulse of ODDM modulation $u(t)$, referred to as the DDOP.\footnote{To ensure that the baseband ODDM {waveform} occupies a symmetric spectrum around \( f = 0 \), the digital symbols \( X_{\textrm{dD}}[m,n] \) are modulated onto both positive and negative carrier frequencies, instead of being restricted only to positive carrier frequencies. {This is reflected in \eqref{Equ:xlt_A-ODDM}, where the summation index $n$ ranges from $-\frac{N}{2}$ to $\frac{N}{2}-1$, resulting in carriers ranging from $-\frac{1}{2T}$ to $\frac{1}{2T}-\frac{1}{NT}$.}}$^,$\footnote{{The expression \eqref{Equ:xlt_A-ODDM} can alternatively be written with the summation taken over only positive indices, eliminating the need for mod-$N$ operation—however, this requires the inverse of the mod-$N$ operation. This results in 
$\bar{x}_{m}(t)=u^{\textrm{cp}}_m(t)\sum_{n=0}^{N-1} X_{\mathrm{dD}}\left[m,n\right]e^{j2\pi \psi_{N}(n) \mathcal{F}t}$. 
}}
Finally, $\bar{x}_{m}(t)$, $\forall m\in\mathcal{S}_{0,M}$, are sent through a TMX parameterized by the delay (fine time) resolution of ODDM $\mathcal{T}\triangleq\frac{T}{M}$. In doing so,  the transmit-ready baseband {waveform} is obtained as
\begin{align} \label{Equ:xt_A-ODDM}
x_{\textrm{cp}}^{\textrm{A-ODDM}}(t)&=\sum_{m=0}^{M-1} \bar{x}_{m}\left(t-m\mathcal{T}\right
).
\end{align}


As for $u^{\textrm{cp}}_m(t)$ in~\eqref{Equ:xlt_A-ODDM}, it is obtained based on the DDOP. 
The DDOP is {a pulse train composed of} $N$ sub-pulses, $\tilde{a}(t)$, that are spaced apart by the duration $T$, leading to~\cite{2022_TWC_JH_ODDM}
\begin{align}\label{Equ:ut}
u(t)=\frac{1}{\sqrt{N}}\sum_{\dot{n}=0}^{N-1}\tilde{a}\left(t-\dot{n}T\right).
\end{align}
The sub-pulse $\tilde{a}(t)$ is the truncated version of an SRN pulse that is parameterized by the zero inter-symbol-interference (ISI) interval which is equal to $\mathcal{T}\triangleq\frac{T}{M}$, the duration $T_{a}\triangleq2Q\frac{T}{M}\ll T$, where $Q$ is a positive integer 
, and~excess bandwidth usage $\beta\frac{M}{T}$, where $0\leqslant\beta\leqslant1$~\cite{2022_TWC_JH_ODDM}. 
{Based on~\eqref{Equ:ut}, the
CP-appended version of DDOP 
$u^{\textrm{cp}}_m(t)$}
  is obtained as $u^{\textrm{cp}}_m(t)=u(t)$, if $0\leqslant m\leqslant M-L_{\textrm{cp}}-1$, and 
\begin{align}\label{Equ:utcp}
u^{\textrm{cp}}_m(t)&=\begin{cases}
u(t+NT),&-T-\frac{T_{a}}{2}\leqslant t < -\frac{T_{a}}{2}, \\
u(t),&\textrm{elsewhere}.
\end{cases}
\notag\\&=\frac{1}{\sqrt{N}} \sum_{\dot{n}=-1}^{N-1}\tilde{a}\left(t-\dot{n}T\right),
\end{align}
if $M-L_{\textrm{cp}}\leqslant m\leqslant M-1$, {where the CP length $T_{\textrm{cp}}$ is $T_{\textrm{cp}}\triangleq L_{\textrm{cp}}\mathcal{T}>\tau_{\mathrm{max}}$, $L_{\textrm{cp}}\in \mathbb{Z}^+$,  and $\tau_{\mathrm{max}}$ is the delay spread of the channel.}
{
We clarify that the use of CP enables ODDM to make the channel behave like a semi-circular convolution (aside from Doppler shift), thereby avoiding ISI. }

\subsubsection{Receiver}


{As shown} in Fig.~\ref{Fig:A-ODDM}(b), $r(t)$ is first truncated based on the DDOP $u(t)$ to obtain 
\begin{align}\label{Equ:yl_A-ODDM}
y_m(t)= r(t)u^*(t-m\mathcal{T}),~~~~~~~~\forall m\in\mathcal{S}_{0,M}. 
\end{align}
Thereafter, $y_m(t)$ is sent through mixers associated with the carriers $n \mathcal{F}$, where $n\in\mathcal{S}_{\frac{N}{2},\frac{N}{2}}$, followed by accumulators
, and finally samplers at $t=NT+T_{a}$. In doing so, the received DD domain symbols $Y_{\textrm{dD}}[m,n]$ 
are obtained as 
\begin{align}\label{Equ:YDD_A-ODDM}
Y_{\textrm{dD}}\left[m,n\right]&=\int r(\bar{t})u^*(\bar{t}-m\mathcal{T})e^{-j2\pi \psi_N(n)\mathcal{F}(\bar{t}-m\mathcal{T})}\mathrm{d}\bar{t}, \notag\\
&~~~~~~~~~~~~~~~~~~~~\forall m\in\mathcal{S}_{0,M}, n\in\mathcal{S}_{0,N}. 
\end{align}
Finally, the symbol detection and channel estimation can be performed based on $Y_{\textrm{dD}}[m,n]$.

It has to be noted that {analog ODDM transmitters} necessitate $MN$ mixers, $M$ windows, and a TMX. Moreover, {the analog ODDM receivers} necessitate  $MN$ mixers, $M$ windows, $MN$ integrators, and $MN$ samplers. These extensive requirements, in particular the $2MN$ mixers, result in prohibitively high implementation complexity for analog ODDM systems.

\subsection{{Approximate} Digital ODDM System}

By leveraging the relationship between MC modulation/demodulation and the IDFT/DFT~\cite{1971_Weinstein_OFDM_usingDFT}, \color{black}and introducing an approximation to eliminate the need for $u(t)$-based truncation/windowing and time multiplexing,  the low complexity {approximate digital} implementation of ODDM system was proposed in~\cite[8]{2022_TWC_JH_ODDM}. The corresponding system is shown in Fig. \ref{Fig:D-ODDM}, and will be detailed next. 

\color{black}



\subsubsection{Transmitter}

\begin{figure*}[t]
\centering
\includegraphics[width=1.8\columnwidth]{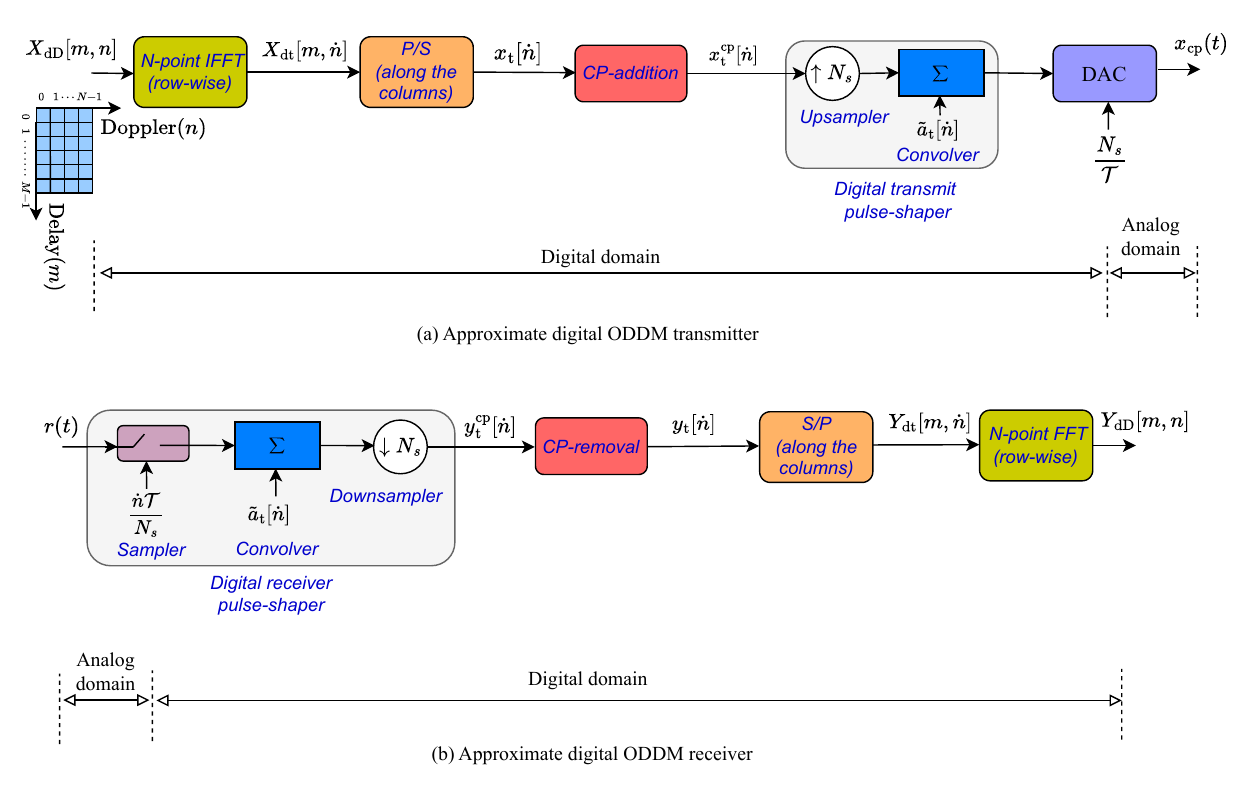}
\vspace{-2mm}
\caption{Illustrations of the transmitter and the receiver of the {approximate digital} ODDM system.}\label{Fig:D-ODDM} 
\end{figure*}

The {approximate digital} {ODDM transmitter conducts} all the transformations and pulse shaping in the digital domain, and then uses a high-rate digital-to-analog converter (DAC) to obtain the transmit-ready {waveform}. 
In particular, as shown in Fig.~\ref{Fig:D-ODDM}(a),  first, row-wise $N$-point IDFT is performed on $X_{\textrm{dD}}[m,n]$ to obtain the delay-time domain symbols $X_{\textrm{\textrm{d}t}}[m,\dot{n}]$ as 
\begin{align} \label{Equ:XDT_D-ODDM}
X_{\textrm{{\textrm{d}t}}}[m,\dot{n}]&=\frac{1}{\sqrt{N}}
\sum_{n=0}^{N-1} X_{\textrm{dD}}[m,n
] e^{j2\pi  \frac{n\dot{n}}{N}},\notag\\ &~~~~~~~~~~~~~~~~~~~~~~~~~\forall m\in\mathcal{S}_{0,M},\dot{n}\in\mathcal{S}_{0,N},
\end{align}
where $\dot{n}$ denotes the time index.
Then, $X_{\textrm{{\textrm{d}t}}}[m,\dot{n}]$ is serialized along the columns to obtain the discrete time domain symbols $x_{\textrm{t}}
[\dot{n}]$, $\forall \dot{n}\in\mathcal{S}_{0,MN}$, with 
$x_{\textrm{t}}
[\bar{\dot{n}}M+m]=X_{\textrm{{\textrm{d}t}}}[m,\bar{\dot{n}}], \forall m\in\mathcal{S}_{0,M},\bar{\dot{n}}\in\mathcal{S}_{0,N}$.
Next, $x_{\textrm{t}}
[\dot{n}]$ is appended with a CP of $L_{\textrm{cp}}$ samples 
to obtain 
\begin{align} \label{Equ:xtcpn_D-ODDM}
x_{\textrm{t}}^{\mathrm{cp}}[\dot{n}]=x_{\textrm{t}}[[\dot{n}]_{MN}], ~~~~~~~~~~~~~\forall \dot{n}\in\mathcal{S}_{L_{\textrm{cp}},MN}.
\end{align}
Finally, $\tilde{a}(t)$-based digital transmit pulse shaping on $x_{\textrm{t}}^{\mathrm{cp}}[\dot{n}]$ and then digital-to-analog conversion based on the the rate $\frac{N_s}{\mathcal{T}}$ are performed to obtain the transmit-ready baseband {waveform} 
\begin{align} \label{Equ:xt_D-ODDM}
x^{\textrm{D-ODDM}}_{\mathrm{cp}}(t)&=\sum_{\dot{n}=-L_{\textrm{cp}}}^{MN-1}x_{\textrm{t}}^{\mathrm{cp}}[\dot{n}]
\tilde{a}(t-\dot{n}\mathcal{T}).
\end{align}
We clarify that as shown in Fig.~\ref{Fig:D-ODDM}(a), the digital transmit pulse shaping involves~upsampling $x_{\textrm{t}}^{\mathrm{cp}}[\dot{n}]$ with an upsampling factor $N_s$ and performing convolution operation 
digitally on the discretized version of $\tilde{a}(t)$, given by $\tilde{a}_{\textrm{t}}[\dot{n}]\triangleq\tilde{a}(t)|_{t=\dot{n}\frac{\mathcal{T}}{N_s}}$.

\color{black}
\begin{remark}\label{Rem:HUP}
We note that, similar to OFDM, if the digital implementation of ODDM was based solely on leveraging the relationship between MC modulation/demodulation and the IDFT/DFT, the resulting transmitter would be as in Fig. \ref{Fig:A-ODDM_1} in the Appendix A. However, the digital implementation of ODDM considered in this work in Fig. \ref{Fig:D-ODDM}—originally introduced in\cite{2022_TWC_JH_ODDM}—incorporates an additional approximation to reduce the implementation complexity associated with truncation/windowing and time multiplexing based on the DDOP function $u(t)$ which has a relatively long duration. In particular, it involves enabling the DAC with the rate of ${N}\mathcal F=\frac{1}{T}$ and the $u(t)$-based truncation/windowing operations to be \emph{approximated} with $a(t)$-based filtering/sample-wise pulse-shaping. This approximation introduces differences in the spectral and orthogonality characteristics between the analog and {approximate digital} waveforms, which will be discussed in detail in the following sections. 
Appendix~A provides a detailed explanation of the steps involved in obtaining the {approximate digital} ODDM implementation considered in this work, in comparison to the analog ODDM implementation and other digital implementation variants.
\end{remark}
\color{black}

\subsubsection{Receiver}
In the {approximate digital} ODDM receiver,
as shown in Fig.~\ref{Fig:D-ODDM}(b), first $\tilde{a}(t)$-based digital receiver pulse shaping (matched filtering) is performed on the received {waveform} $r(t)$ to obtain
\begin{align}\label{Equ:yt_D-ODDM}
y_{\textrm{t}}^{\mathrm{cp}}[\dot{n}]=\int r(\bar{t})a^*(\bar{t}-\dot{n}\mathcal{T})\mathrm{d}\bar{t}, ~~~~~~~~~~\forall \dot{n}\in\mathcal{S}_{D,MN}. 
\end{align}
We clarify that the digital receiver pulse shaping involves (i)~sampling $r(t)$ at the interval of $\frac{\mathcal{T}}{N_s}$, (ii) performing convolution operation digitally based on $\tilde{a}^{*}_{\textrm{t}}[\dot{n}]$, and (iii) downsampling the resulting {waveform} based on the downsampling factor $N_s$.

Next, the symbols corresponding to the CP in $y_{\textrm{t}}^{\mathrm{cp}}[\dot{n}]$ (i.e., the first $L_{\textrm{cp}}$
symbols) are removed to obtain 
\begin{align} \label{Equ:ytn_D-ODDM}
y_{\textrm{t}}[\dot{n}]=y_{\textrm{t}}^{\mathrm{cp}}[\dot{n}], ~~~~~~~~~~~~~\forall \dot{n}\in\mathcal{S}_{0,MN}.
\end{align}
Thereafter, $y_{\textrm{t}}[\dot{n}]$ is rearranged 
to obtain discrete delay time
domain symbols as
\begin{align} \label{Equ:Ydt_D-ODDM}
Y_{\textrm{\textrm{d}t}}[m,\dot{n}]=y_{\textrm{t}}
[nM+m],  ~~~~~\forall m\in\mathcal{S}_{0,M},\dot{n}\in\mathcal{S}_{0,N}.
\end{align}
Then, row-wise $N$-point DFT is performed on $Y_{\textrm{{\textrm{d}t}}}[m,\dot{n}]$ to obtain the received DD domain symbols as 
\begin{align} \label{Equ:YDD_D-ODDM}
Y_{\textrm{{dD}}}[m,n]&=\frac{1}{\sqrt{N}}
\sum_{\dot{n}=0}^{N-1} Y_{\textrm{\textrm{d}t}}[m,\dot{n}
] e^{-j2\pi  \frac{n\dot{n}}{N}}, \notag\\
&~~~~~~~~~~~~~~~~~~~~~~~\forall m\in\mathcal{S}_{0,M},n\in\mathcal{S}_{0,N}.
\end{align}

We note that the {{approximate digital} ODDM transmitter} involves numerous digital operations and the utilization of only a single high-frequency DAC. Furthermore, {{approximate digital} ODDM receivers} require a sampler and the execution of several digital operations. These digital operations necessitate complex-valued multiplications and memory blocks. 

\section{Time and Frequency Domain Representations of {Analog and Approximate Digital ODDM Waveforms}} 
\label{Sec:TDandFD}



Differences in analog and {approximate digital} ODDM systems lead to variations in their transmit waveforms. 
{In this section, 
we first start by investigating the {time and frequency domain/Fourier representations} of the basis functions of their waveforms. Thereafter, we delve into the power spectral density analysis.}

\subsection{Time Domain Representation {of Basis Functions}}


We first represent the CP-less transmit waveforms 
using their basis functions $\phi_{m,n}^{\Xi\textrm{-ODDM}}(t)$ as
\begin{align} \label{Equ:xt_ODDM}
x^{\Xi\textrm{-ODDM}}(t)&=\sum_{m=0}^{M-1}\sum_{n=0}^{N-1} X_{\mathrm{dD}}[m,n]\phi_{m,n}^{\Xi\textrm{-ODDM}}(t),
\end{align}
where $\Xi\in\{\mathrm{A},\mathrm{D}\}$. \textcolor{black}{Here, the basis function $\phi_{m,n}^{\Xi\textrm{-ODDM}}(t)$ can be viewed as the {waveform} that carries the $(m,n)$-th symbol in $X_{\mathrm{dD}}[m,n]$.}

\begin{figure}[t]
\centering
\includegraphics[width=1\columnwidth]{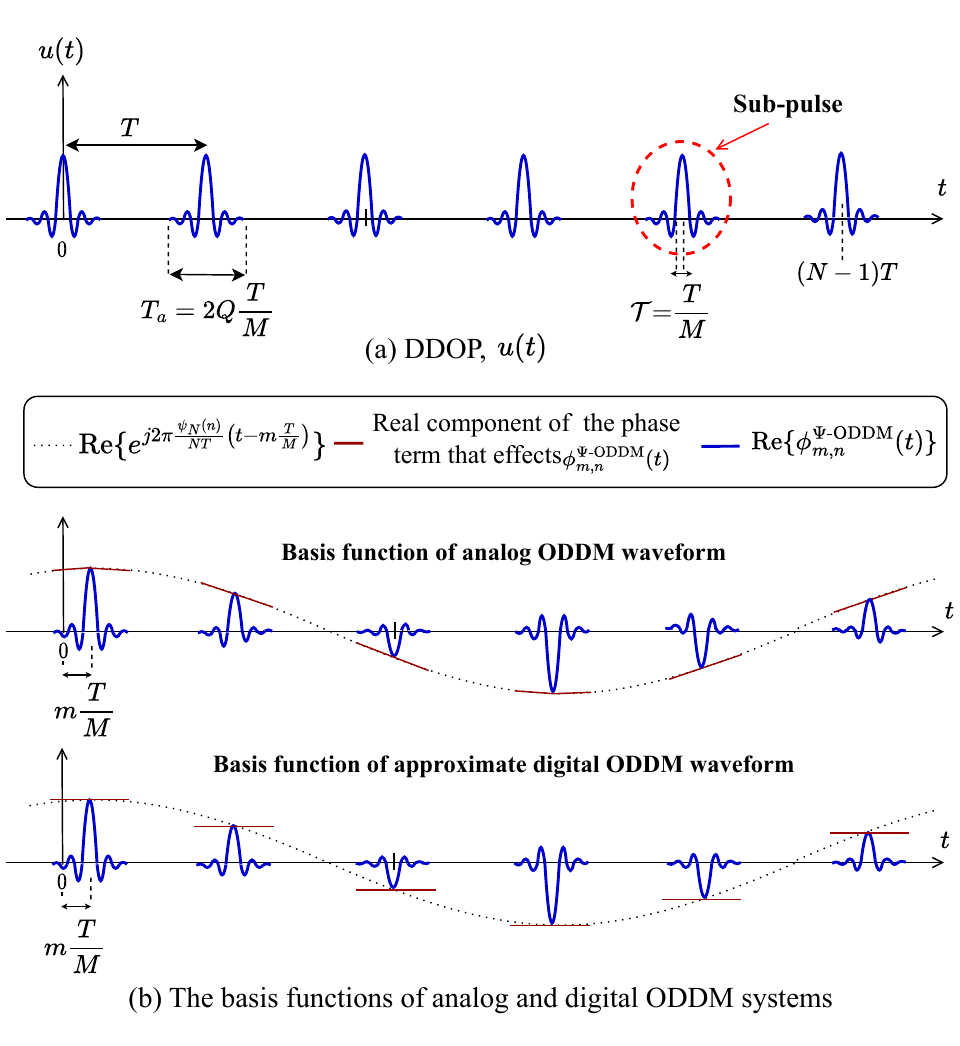}
\caption{Illustration of (a) the DDOP $u(t)$ and (b) the basis functions of analog and digital ODDM systems (consider $N=6$ and $n=1$).}\label{Fig:phi_lk}
\end{figure}

Based on~\eqref{Equ:xlt_A-ODDM}-\eqref{Equ:ut}, the basis function of analog ODDM waveforms can be obtained as
\begin{align} \label{Equ:phi_A-ODDM}
&\phi_{m,n}^{\textrm{A-ODDM}}(t)\notag\\
&~~= e^{j2\pi  \frac{\psi_N(n)}{NT}\left(t-m\frac{T}{M}\right
)}u\left(t-m\frac{T}{M}\right
)\notag\\
&~~=\frac{1}{\sqrt{N}}\sum_{\dot{n}=0}^{N-1}e^{j2\pi  \frac{\psi_N(n)}{NT}\left(t-m\frac{T}{M}\right
)} \tilde{a}\left(t-\dot{n}T-m\frac{T}{M}\right
).
\end{align}
\noindent
Based on \eqref{Equ:XDT_D-ODDM}-\eqref{Equ:xt_D-ODDM}, the basis functions of {approximate digital} ODDM waveforms can be obtained as
\begin{align} \label{Equ:phi_D-ODDM}
&\phi_{m,n}^{\textrm{D-ODDM}}(t)\notag\\
&=\!\frac{1}{\sqrt{N}}\!\!\sum_{\dot{n}=0}^{N-1}e^{j2\pi \frac{n\dot{n}}{N}}\tilde{a}\left(t-\dot{n}T-m\frac{T}{M}\right)\notag\\
&=\!\frac{1}{\sqrt{N}}\!\!\sum_{\dot{n}=0}^{N-1}e^{j2\pi  \frac{\psi_N(n)}{NT}\left(t-m\frac{T}{M}\right
)}|_{t=\dot{n}T+m\frac{T}{M}}\tilde{a}\left(t-\dot{n}T-m\frac{T}{M}\right)\!.
\end{align}
A schematic illustration of $\phi_{m,n}^{\textrm{A-ODDM}}(t)$ and $\phi_{m,n}^{\textrm{D-ODDM}}(t)$ are given in Fig.~\ref{Fig:phi_lk}(b).

Based on~\eqref{Equ:phi_A-ODDM} and \eqref{Equ:phi_D-ODDM}, it is interesting to 
observe that the basis functions of both analog and {approximate digital} ODDM waveforms are defined by a sub-pulse train and a phase term. The sub-pulse train is the same for both and can be represented as the delayed version of $u(t)$.
{However, the phase term differs.} Particularly,   as shown in Fig.~\ref{Fig:phi_lk}(b), the phase term varies over the duration of the basis functions for both. However, \textit{while a time-varying phase term is experienced by a specific sub-pulse in the basis function of analog ODDM waveforms, a fixed phase term is experienced by the same sub-pulse in the basis function of {approximate digital} ODDM waveforms} \cite{2022_TWC_JH_ODDM}. 


We clarify that the phase term difference observed between the analog and {approximate digital} waveforms stems from the distinct ways in which phase terms are applied to \( X_{\mathrm{dD}}[m,n] \) in their respective systems. In the analog ODDM system, as depicted in Fig. \ref{Fig:A-ODDM}, a time-varying phase term \( e^{j2\pi \frac{\psi_N(n)}{NT}\left(t-m\frac{T}{M}\right)} \) is introduced to \( X_{\mathrm{dD}}[m,n] \) using mixers. On the other hand, in the {approximate digital} ODDM system, as shown in Fig. \ref{Fig:D-ODDM}, the sampled version of the same time-varying phase term, \( e^{j2\pi \frac{\psi_N(n)}{NT}\left(t-m\frac{T}{M}\right)}|_{t=\dot{n}T+m\frac{T}{M}} \),  at discrete time instances \( t = \dot{n}T + m\frac{T}{M} \) is introduced to \( X_{\mathrm{dD}}[m,n] \) using the IFFT operation, where \( \dot{n} \in \mathcal{S}_{0,N} \).\footnote{{
The highlighted phase difference observed between the analog and digital ODDM waveforms is not an inherent characteristic of all digital implementations, but rather a consequence of the approximation in the considered {approximate digital} implementation of ODDM. For instance, the direct digital implementation shown in Fig.~\ref{Fig:A-ODDM_1} reproduces the waveform of the analog approach without any phase discrepancy.
}}

\begin{figure}[t]
\centering
%
\includegraphics[width=1\columnwidth]{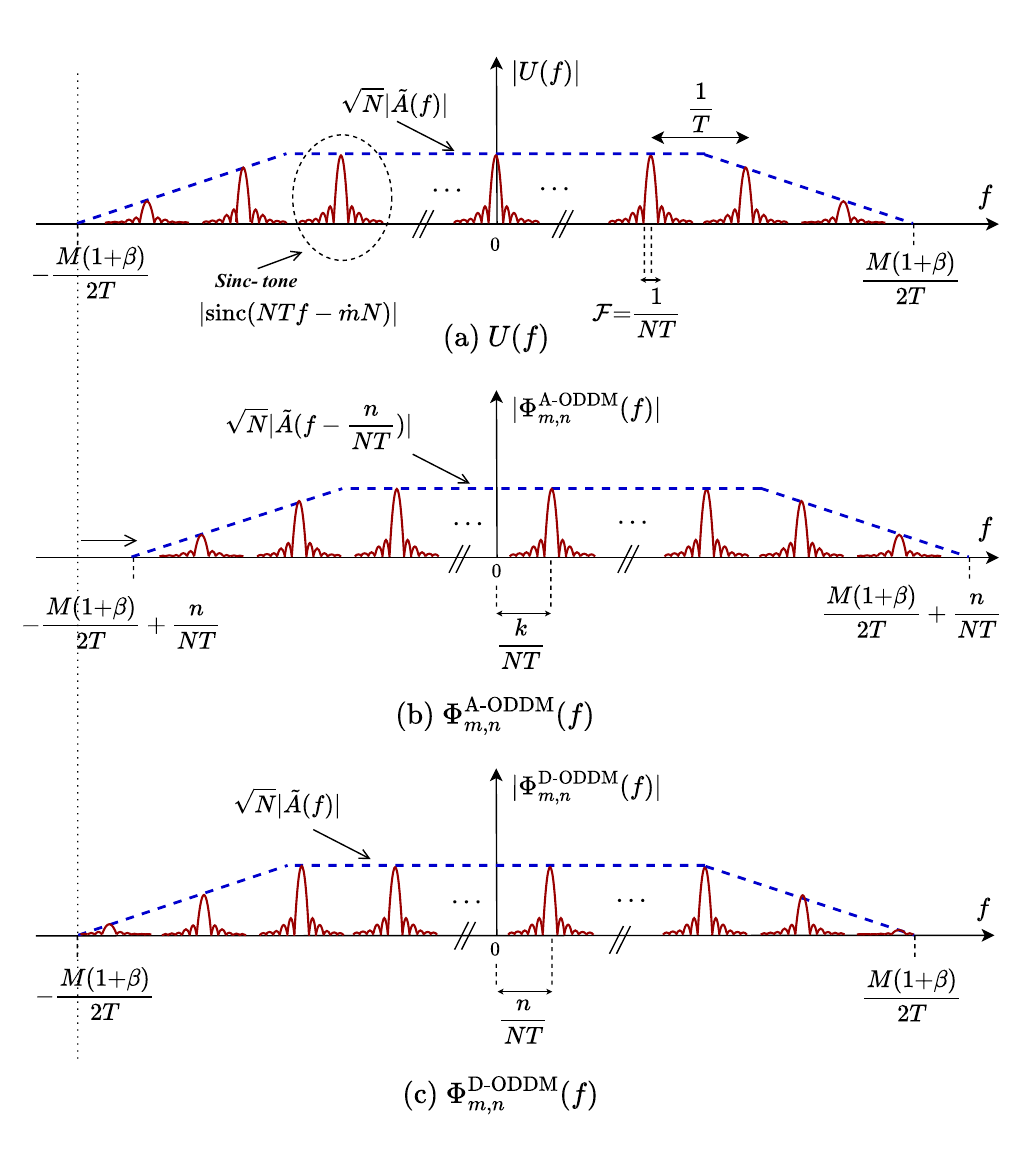}
\caption{Illustration of the {Fourier/frequency domain representations} of the DDOP and the basis functions of analog and digital ODDM systems (consider $0<n<N/2$).}\label{Fig:Phi_lk}
\end{figure}
Despite the differences in the basis functions of analog and {approximate digital} ODDM waveforms, as shown in \cite[Fig. 20]{2023_ODDM_TCOM}, the difference between their transmit waveforms is marginal. This is because the energy of the sub-pulses in the basis functions is primarily concentrated within the duration captured by their initial several zero crossings. Within this duration, the phase term $e^{j2\pi \frac{\psi_N(n)}{NT}(t-m\frac{T}{M})}$ in the basis function of analog ODDM waveforms can be approximated to the phase term $e^{j2\pi \frac{n\dot{n}}{N}}$ in {approximate digital} ODDM waveforms. 

\begin{figure*}[b]
\hrulefill
\normalsize 
\begin{align} 
\label{Equ:Uf}
U(f)&=\sqrt{N} \tilde{A}(f)\sum_{\dot{m}=-\infty}^{\infty}e^{-j \pi (N-1)\left(Tf-\dot{m}\right)}\sinc(NTf-\dot{m}N)\\
\label{Equ:Phi_A-ODDM}
\Phi_{m,n}^{\textrm{A-ODDM}}(f)&= \underbrace{\sqrt{N} e^{-j2\pi  fm\frac{T}{M}} \tilde{A}\left(f-\frac{\psi_N(n)}{NT}\right) }_{\substack{\textrm{Envelope}}} \underbrace{ \sum_{\dot{m}=-\infty}^{\infty}e^{-j \pi (N-1)\left(T(f-\frac{\psi_N(n)}{NT})-\dot{m}\right)}\sinc(NTf-\dot{m}N-\psi_N(n))}_{\substack{\textrm{Infinite train of sinc tones}}} \\
\label{Equ:Phi_D-ODDM}
\Phi_{m,n}^{\textrm{D-ODDM}}(f)
&= \underbrace{ \sqrt{N} e^{-j2\pi  fm\frac{T}{M}}\tilde{A}(f)   }_{\substack{\textrm{Envelope}}} \underbrace{\sum_{\dot{m}=-\infty}^{\infty} e^{-j \pi (N-1)\left(T(f-\frac{\psi_N(n)}{NT})-\dot{m}\right)}\sinc(NTf{-}\dot{m}N{-}\psi_N(n))}_{\substack{\textrm{Infinite train of sinc tones}}}.
\end{align}
\end{figure*}


\subsection{Frequency Domain Representation  {of Basis Functions}}



The {frequency domain/Fourier representation} of the transmit waveforms 
can be represented as 
\begin{align} \label{Equ:Xf_ODDM}
X^{\Xi\textrm{-ODDM}}(f)&=\sum_{m=0}^{M-1}\sum_{n=0}^{N-1} X_{\mathrm{dD}}[m,n]\Phi_{m,n}^{\Xi\textrm{-ODDM}}(f),
\end{align} 
where $\Phi_{m,n}^{\Xi\textrm{-ODDM}}(f)$ is the {frequency domain representation} of the basis functions $\phi_{m,n}^{\Xi\textrm{-ODDM}}(t)$.
The $\Phi_{m,n}^{\Xi\textrm{-ODDM}}(f)$ for analog and {approximate digital} ODDM waveforms are presented next in the following lemma. 



\begin{lemma}\label{Lemma:Phi_ODDM}
The {frequency domain representations} of the basis functions of analog ODDM {waveform} $\Phi_{m,n}^{\textrm{A-ODDM}}(f)$ and {approximate digital} ODDM waveforms  $\Phi_{m,n}^{\textrm{D-ODDM}}(f)$ are given by
\eqref{Equ:Phi_A-ODDM} and \eqref{Equ:Phi_D-ODDM}, respectively. The expressions~\eqref{Equ:Uf}-\eqref{Equ:Phi_D-ODDM} are given at the end of this page.  Therein, $U(f)$ and $\tilde{A}(f)$
are the {frequency domain representations} of $u(t)$ and $\tilde{a}(t)$, respectively.

\textit{Proof:} The proof is given in Appendix A.
\hfill $\blacksquare$
\end{lemma}

A schematic illustration of $U(f)$, $\Phi_{m,n}^{\textrm{A-ODDM}}(f)$, and $\Phi_{m,n}^{\textrm{D-ODDM}}(f)$ is given in Fig.~\ref{Fig:Phi_lk}.

Based on~\eqref{Equ:Phi_A-ODDM}-\eqref{Equ:Phi_D-ODDM} in Lemma~\ref{Lemma:Phi_ODDM} and Fig.~\ref{Fig:Phi_lk},
we observe that $\Phi_{m,n}^{\textrm{A-ODDM}}(f)$ and $\Phi_{m,n}^{\textrm{D-ODDM}}(f)$ can be represented by using an \textit{infinite train of sinc tones} and an \textit{envelope function}. The infinite train of sinc tones is the same in both $\Phi_{m,n}^{\textrm{A-ODDM}}(f)$ and $\Phi_{m,n}^{\textrm{D-ODDM}}(f)$, and can be obtained by frequency-shifting the infinite train of sinc tones that appears in $U(f)$. However, it is interesting to 
observe that
\textit{while the envelope of $\Phi_{m,n}^{\textrm{A-ODDM}}(f)$ is influenced by the frequency shifted version of $\tilde{A}(f)$ (i.e.,  $\tilde{A}(f-\frac{\psi_N(n)}{NT})$), 
the envelope of $\Phi_{m,n}^{\textrm{D-ODDM}}(f)$ is directly influenced by $\tilde{A}(f)$.} Consequently, if we disregard the side-lobes in $\tilde{A}(f)$ caused by  the truncation of sub-pulses $\tilde{a}(t)$ in the time domain, it follows that, 
on one hand, the basis functions of analog ODDM waveforms $\phi_{m,n}^{\textrm{A-ODDM}}(t)$ occupies the spectrum spanning from
 ${-}\frac{M(1{+}\beta)}{2T}+\frac{\psi_N(n)}{NT}$ to $\frac{M(1{+}\beta)}{2T}+\frac{\psi_N(n)}{NT}$  (see Fig.~\ref{Fig:Phi_lk}(b)).\footnote{Given that $\tilde{a}(t)$ is a truncating pulse, $\tilde{a}(t)=a(t)\times \Pi_{T_a}\left(t\right)$, where $a(t)$ is the untruncated SRN pulse. Hence, $\tilde{A}(f)$ in~\eqref{Equ:Uf}-\eqref{Equ:Phi_D-ODDM} is $\tilde{A}(f)=A(f)\star\sinc (T_af)$, where $A(f)$ is the {frequency domain representation} of $a(t)$ and $\star $ denotes the linear convolution operation. Thus, $\tilde{A}(f)$  experiences side-lobes beyond the spectrum of $A(f)$ spanning  from ${-}\frac{M(1{+}\beta)}{2T}$ to $\frac{M(1{+}\beta)}{2T}$, leading to OOBE.} On the other hand, the basis functions of {approximate digital} ODDM waveforms $\phi_{m,n}^{\textrm{D-ODDM}}(t)$ occupy the spectrum spanning from 
 ${-}\frac{M(1{+}\beta)}{2T}$ to $\frac{M(1{+}\beta)}{2T}$ (see Fig.~\ref{Fig:Phi_lk}(c)).

\subsection{{Power Spectral Density Analysis}}

{We next derive the expected power spectral densities, the results of which are provided in the 
following theorem. }

\begin{theorem}\label{Thr:EPS_ODDM}
Suppose that the information-bearing symbols $X_{\textrm{dD}}[m,n]$ are i.i.d satisfying $E[|X_{\textrm{dD}}[m,n]|^2] = E_s$ and $E[X_{\textrm{dD}}[m,n]] = 0$, where $E_s$ is the energy per symbol in $X_{\textrm{dD}}[m,n]$.
The expected  power spectral densities of analog and {approximate digital} ODDM transmit waveforms are 
\begin{align} \label{Eq:EPSD_A-ODDM}
&\mathbb{E}\left[\left|X^{\textrm{A-ODDM}}(f)\right|^2\right]\\
&\!\!= E_sNM  \!\!\sum_{n={-}\frac{N}{2}}^{\frac{N}{2}{-}1}\left|\tilde{A}\left(f{-}\frac{n}{NT}\right)\right|^2\underbrace{\sum_{\dot{m}={-}\infty}^{\infty}\sinc^2(NTf{-}mN{-}n),}_{\substack{\textrm{Infinite train of sinc{-}squared tones}}}\notag
\end{align}

\noindent
and
\begin{align} 
\label{Eq:EPSD_D-ODDM}
&\mathbb{E}\left[\left|X^{\textrm{D-ODDM}}(f)\right|^2\right]\notag\\
&~~~~~= E_sNM  \left|\tilde{A}(f)\right|^2\underbrace{\sum_{\dot{m}=-\infty}^{\infty}\sinc^2(NTf-\dot{m})}_{\substack{\textrm{Dense infinite train of sinc-squared tones}}},
\end{align}
respectively.

\textit{Proof:} The proof is given in Appendix B.
\hfill $\blacksquare$
\end{theorem}

\begin{figure*}[h]
\centering
\includegraphics[width=1.9\columnwidth]{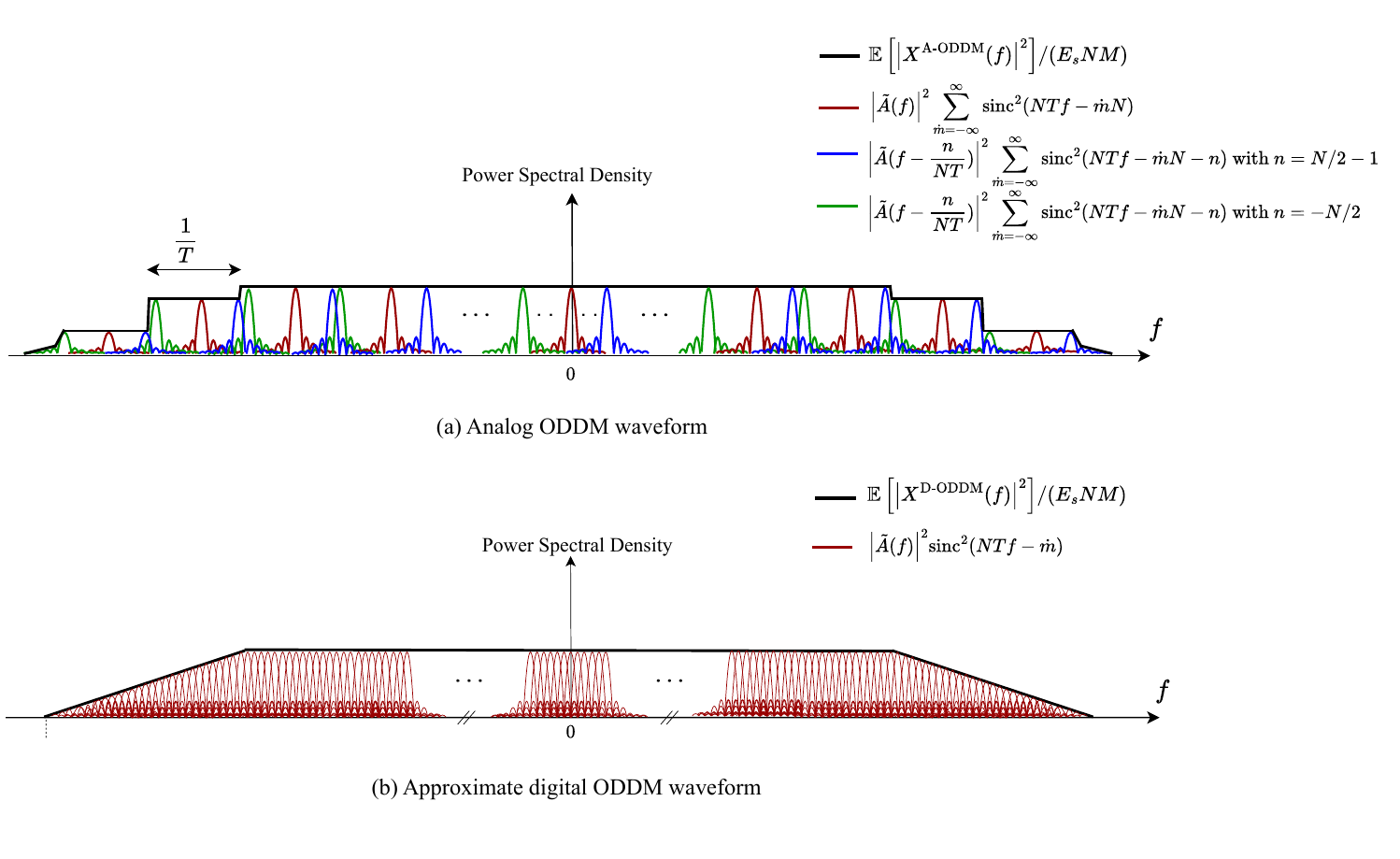}
\vspace{-4.5mm}
\caption{Schematic illustration of the expected power spectral densities of the waveforms in analog and digital ODDM systems.}\label{Fig:EPSD} 
\end{figure*}

A schematic illustration of {$\mathbb{E}\left[\left|X^{\textrm{A-ODDM}}(f)\right|^2\right]$ and $\mathbb{E}\left[\left|X^{\textrm{D-ODDM}}(f)\right|^2\right]$} 
is given in Fig.~\ref{Fig:EPSD}.

Based on~\eqref{Eq:EPSD_A-ODDM}, we deduce that multiple replicas of an infinite train of sinc-squared tones with an envelope $|\tilde{A}(f)|^2$ (i.e., $\left|\tilde{A}(f)\right|^2\sum_{\dot{m}=-\infty}^{\infty}\sinc^2(NTf-\dot{m}N)$) are frequency-shifted by different amounts  (i.e., $\frac{n}{NT}$, where $n\in \mathcal{S}_{\frac{N}{2},\frac{N}{2}}$) and then aggregated together to form $\mathbb{E}\left[\left|X^{\textrm{A-ODDM}}(f)\right|^2\right]$. This process of shifting the same train of tones by different amounts and then aggregating them leads to, as shown in Fig.~\ref{Fig:EPSD}(a), \textit{the spectrum of analog ODDM {waveform} to exhibit 
a step-wise behavior in its transition regions with a step width of $1/T$.} 

Moreover, based on~\eqref{Eq:EPSD_D-ODDM}, we deduce that a dense infinite train of sinc-squared tones 
with an envelope $|\tilde{A}(f)|^2$ (i.e., $\left|\tilde{A}(f)\right|^2\sum_{\dot{m}=-\infty}^{\infty}\sinc^2(NTf-\dot{m})$) forms $\mathbb{E}\left[\left|X^{\textrm{D-ODDM}}(f)\right|^2\right]$.
Consequently, as shown in Fig.~\ref{Fig:EPSD}(b), \textit{the {spectrum} of {approximate digital} ODDM waveforms are confined to the 
{spectrum} of the sub-pulse $\left|\tilde{A}(f)\right|^2$}. Additionally,  unlike analog ODDM systems,  the {spectrum} of {approximate digital} ODDM waveforms does not display any step-wise behavior. 





\section{Orthogonality Characteristics of Analog and Digital ODDM waveforms}
\label{Sec:Orthogonality}

In this section, we investigate the orthogonality characteristics of analog and {approximate digital} ODDM systems. 
We first summarize the result related to orthogonality characteristics of the analog  ODDM {waveform} that were presented in~\cite{2022_TWC_JH_ODDM,2023_ODDM_TCOM}. Thereafter, we derive the orthogonality characteristics of the {approximate digital} ODDM waveforms. 

\subsection{Analog ODDM Systems}

The orthogonality characteristics of analog ODDM waveforms were investigated in~\cite{2022_ICC_JH_ODDM,2022_TWC_JH_ODDM,2022_GC_JH_ODDMPulse,2023_ODDM_TCOM}.
In particular,~\cite{2022_ICC_JH_ODDM,2022_TWC_JH_ODDM} showed that the ambiguity function of $u(t)$, given by $A_{u,u}(t,f)\triangleq\int u(\bar{t})u^*(\bar{t}-t)e^{-j2\pi f(\bar{t}-t)}\mathrm{d}\bar{t}$,  satisfies 
\begin{align}\label{Equ:Auu_A-ODDM}
A_{u,u}(\bar{m}\mathcal{T},\bar{n}\mathcal{F})&=\int u(\bar{t})u^*(\bar{t}-\bar{m}\mathcal{T})e^{-j2\pi \bar{n}\mathcal{F}(\bar{t}-\bar{m}\mathcal{T})}\mathrm{d}\bar{t}, \notag\\
&=\delta (\bar{m})\delta (\bar{n}), \notag\\
&~~~~~~~\forall \bar{m}\in\mathcal{S}_{M-1,M}, \bar{n}\in\mathcal{S}_{N-1,N}. 
\end{align} 
From \eqref{Equ:Auu_A-ODDM}, it can be shown that the basis functions of  analog ODDM waveforms form the orthonormal set of\footnote{{The autocorrelation function of the basis functions of classical MC modulation $\phi_{m,n}(t)$, given by $\Psi(m,m',n,n')=\int \phi_{m,n}(t) (\phi_{m',n'}(t))^* \textrm{d}t$, and the ambiguity function of the fundamental/prototype pulse of the MC modulation $\gamma(t)$, given by $A_{\gamma,\gamma}(t,f)=\int_{-\infty}^{\infty} \gamma(\bar{t}) \gamma^*(\bar{t} - t) e^{-j 2 \pi f (\bar{t}-t)} \, d\bar{t}$, are related as $\Psi(m,m',n,n')=e^{j2\pi  n(m'-m)\mathcal{T}\mathcal{F}}A_{\gamma,\gamma}((m'-m)\mathcal{T},(n'-n)\mathcal{F})$.}}
\begin{align}\label{Equ:Orthonorm_A-ODDM}
&\Psi^{\textrm{A-ODDM}}(m,m',n,n')\triangleq\int \phi_{m,n}^{\textrm{A-ODDM}}(t) (\phi_{m',n'}^{\textrm{A-ODDM}}(t) )^* \textrm{d}t \notag\\
&=\delta (m'-m)\delta (n'-n), 
\forall m,m'\in\mathcal{S}_{0,M}, n',n\in\mathcal{S}_{0,N}, 
\end{align} 
thereby leading to analog ODDM waveforms satisfying orthogonality w.r.t. their resolutions $\mathcal{T}\triangleq\frac{T}{M}$ and $\mathcal{F}\triangleq\frac{1}{NT}$.

It has to be noted that the derivations of $A_{u,u}(\bar{m}\mathcal{T},\bar{n}\mathcal{F})$ in \eqref{Equ:Auu_A-ODDM} and $\Psi^{\textrm{A-ODDM}}(m,m',n,n')$ in \eqref{Equ:Orthonorm_A-ODDM} are limited to the scenario where the duration of the sub-pulse $\tilde{a}(t)$ is constrained to   $T_a\ll T$~\cite{2022_TWC_JH_ODDM}. Moreover, they involve two tight approximations, namely, 
$$\textrm{(i)}~\textit{Approx.~1:}~\int_{-T_a/2}^{T_a/2} \tilde{a}(\bar{t})\tilde{a}^*(\bar{t}-\bar{m}\mathcal{T})\mathrm{d}\bar{t}\approx\delta(\bar{m})~~~~~~~~~~~$$ for $\bar{n}=0$ and $\bar{m}$ being either $|\bar{m}|<2Q$ or $|\bar{m}|>M-2Q$ and 



{To validate} the results in \eqref{Equ:Auu_A-ODDM} and \eqref{Equ:Orthonorm_A-ODDM},  we numerically evaluate $A_{u,u}(\bar{m}\mathcal{T},\bar{n}\mathcal{F})$, $\forall \bar{m}\in\mathcal{S}_{M-1,M}, \bar{n}\in\mathcal{S}_{N-1,N}$, and plot the results in Fig.~\ref{Fig:NumOrthAmbuity-A-ODDM}(a). We use the squared root-raised-cosine (SRRC) pulse as the SRN sub-pulse with $\beta=0.15$, and set $T_a\approx 0.3T$. We observe that $A_{u,u}(\bar{m}\mathcal{T},\bar{n}\mathcal{F})$ is 1 for $\bar{m}=\bar{n}=0$. Whereas for $\bar{m}\neq0$ and $\bar{n}\neq0$, the $A_{u,u}(\bar{m}\mathcal{T},\bar{n}\mathcal{F})$ is not necessarily zero for all values of $\bar{m}$ and $\bar{n}$, but is considerably smaller in magnitude compared to 1 (below $-40~\textrm{dB}$). This validates the orthogonality characteristics of analog ODDM waveforms.

\subsubsection{Generalized Analog ODDM Waveforms with Cyclic Prefix and Cyclic Suffix}



To relax the duration constraint $T_a\ll T$
,~\cite{2022_GC_JH_ODDMPulse,2023_ODDM_TCOM} recently proposed a generalized 
analog ODDM {waveform} in which a {CS and an additional CP} \footnote{Please note that this additional CP is different from the CP added to the transmitted {waveform} to combat the delay spread caused by the channel. Refer to~\cite{2022_GC_JH_ODDMPulse,2023_ODDM_TCOM} for more details.} were introduced. In particular, 
it was proposed to utilize a $u_{\textrm{ce}}(t)$-based MC modulator as the transmitter while utilizing a $u(t)$-based MC demodulator at the receiver, where 
\begin{align}\label{ut_general}
u_{\textrm{ce}}(t)=\sum_{\dot{n}=-D}^{N-1+D}\tilde{a}\left(t-(\dot{n}+D)T\right),
\end{align}
is the \emph{generalized DDOP} with its sub-pulse duration set to be 
$T_a\lesseqgtr T$ and $D \triangleq\lceil\frac{T_a}{T}\rceil$ with $\lceil\cdot\rceil$ denoting the ceil operation. 
The cross ambiguity function of $u_{\textrm{ce}}(t)$ and $u(t)$, given by $A_{u_{\textrm{ce}},u}(t,f)\triangleq\int u_{\textrm{ce}}(\bar{t})u^*(\bar{t}-t)e^{-j2\pi f(\bar{t}-t)}\mathrm{d}\bar{t}$,  satisfies 
\begin{align}\label{Equ:Auu_GA-ODDM}
A_{u_{\textrm{ce}},u}(t,f)
&=\delta (\bar{m})\delta (\bar{n}), ~~\forall \bar{m}\in\mathcal{S}_{M-1,M}, \bar{n}\in\mathcal{S}_{N-1,N}. 
\end{align}
Based on \eqref{Equ:Auu_GA-ODDM}, it can be shown that the basis functions of generalized analog ODDM waveforms form another orthonormal set as given in \eqref{Equ:Auu_A-ODDM}.

\begin{figure}[t]
\centering\subfloat[$A_{u,u}(\bar{m}\mathcal{T},\bar{n}\mathcal{F})$ of analog ODDM when $T_a\approx 0.3T$.]{ \includegraphics[width=0.9\columnwidth]{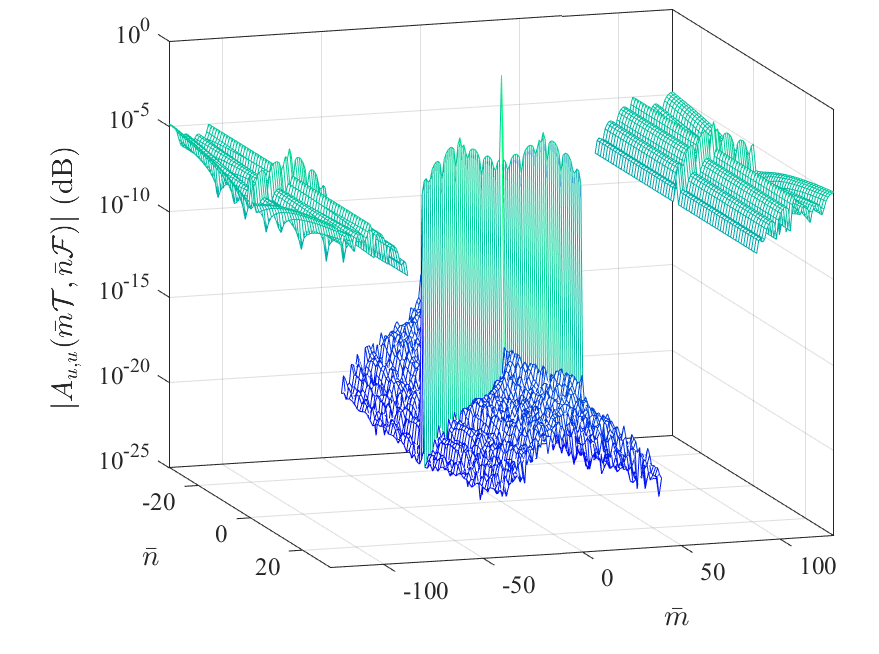}} \hspace{4 mm} 
  \subfloat[$A_{u_{\textrm{ce}},u}(\bar{m}\mathcal{T},\bar{n}\mathcal{F})$ of analog ODDM with {CS and an additional CP} when $T_a=10T$.]{ \includegraphics[width=0.9\columnwidth]{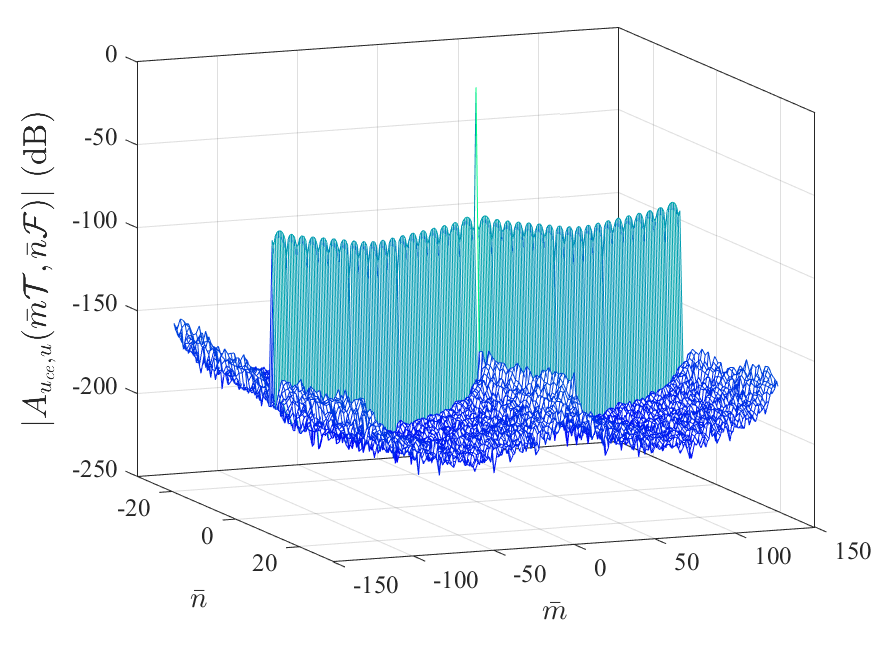}}
  \caption{Illustration of the orthogonality characteristics of analog ODDM systems ($M=128$, $N=32$. The other parameters are same as those used in Section \ref{Sec:Num}.)}\label{Fig:NumOrthAmbuity-A-ODDM}
\end{figure}

It has to be noted that the results for $A_{u_{\textrm{ce}},u}(t,f)$ in \eqref{Equ:Auu_GA-ODDM} 
is not influenced by \textit{Approx.~2} due to the introduction of CS and the additional CP. Moreover, although \textit{Approx.~1} still influences $A_{u_{\textrm{ce}},u}(t,f)$  in \eqref{Equ:Auu_GA-ODDM}, it is more strictly satisfied for generalized analog ODDM systems due to the relaxation of the duration constraint $T_a\ll T$. 

Next, in Fig.~\ref{Fig:NumOrthAmbuity-A-ODDM}(b), 
we numerically evaluate $A_{u_{\textrm{ce}},u}(\bar{m}\mathcal{T},\bar{n}\mathcal{F})$, $\forall \bar{m}\in\mathcal{S}_{M-1,M}, \bar{n}\in\mathcal{S}_{N-1,N}$, for generalized analog ODDM with {CS and an additional CP} while considering $T_a=10T$.
We first observe that similar to that in Fig.~\ref{Fig:NumOrthAmbuity-A-ODDM}(a), $A_{u_{\textrm{ce}},u}(\bar{m}\mathcal{T},\bar{n}\mathcal{F})$ is 1 for $\bar{m}=\bar{n}=0$, and considerably smaller in magnitude compared to 1 (below $-80~\textrm{dB}$) for $\bar{m}\neq0$ and $\bar{n}\neq0$. 
By comparing Fig.~\ref{Fig:NumOrthAmbuity-A-ODDM}(a) with Fig.~\ref{Fig:NumOrthAmbuity-A-ODDM}(b), we observe that, 
on the one hand, for $\bar{n} \neq 0$, while $A_{u,u}(\bar{m}\mathcal{T},\bar{n}\mathcal{F})$ in Fig.~\ref{Fig:NumOrthAmbuity-A-ODDM}(a) was in the order of $-40~\textrm{dB}$ for certain values of $\bar{m}$, $A_{u_{\textrm{ce}},u}(\bar{m}\mathcal{T},\bar{n}\mathcal{F})$ in Fig.~\ref{Fig:NumOrthAmbuity-A-ODDM}(b) is always in the order of $-150~\textrm{dB}$ for all values of $\bar{m}$. This is due to the fact that  $A_{u_{\textrm{ce}},u}(\bar{m}\mathcal{T},\bar{n}\mathcal{F})$ is not influenced by \textit{Approx. 2} due to the introduction of {CS and an additional CP} when $T_a$ is high.
On the other hand,  we observe that for $\bar{m} \neq 0$ and $\bar{n} = 0$, the $A_{u_{\textrm{ce}},u}(\bar{m}\mathcal{T},\bar{n}\mathcal{F})$  in Fig.~\ref{Fig:NumOrthAmbuity-A-ODDM}(b) is lower than $A_{u,u}(\bar{m}\mathcal{T},\bar{n}\mathcal{F})$ in Fig.~\ref{Fig:NumOrthAmbuity-A-ODDM}(a).
This highlights the diminishing effect of \textit{Approx.~1} as $T_a$ increases. 

\subsection{Approximate Digital ODDM Systems}

\begin{figure}[t]
\centering\subfloat[$\Lambda^{\textrm{D-ODDM}}(\bar{m},\bar{n})$ of {approximate digital} ODDM when $T_a\approx 0.3T$.]{ \includegraphics[width=0.9\columnwidth]{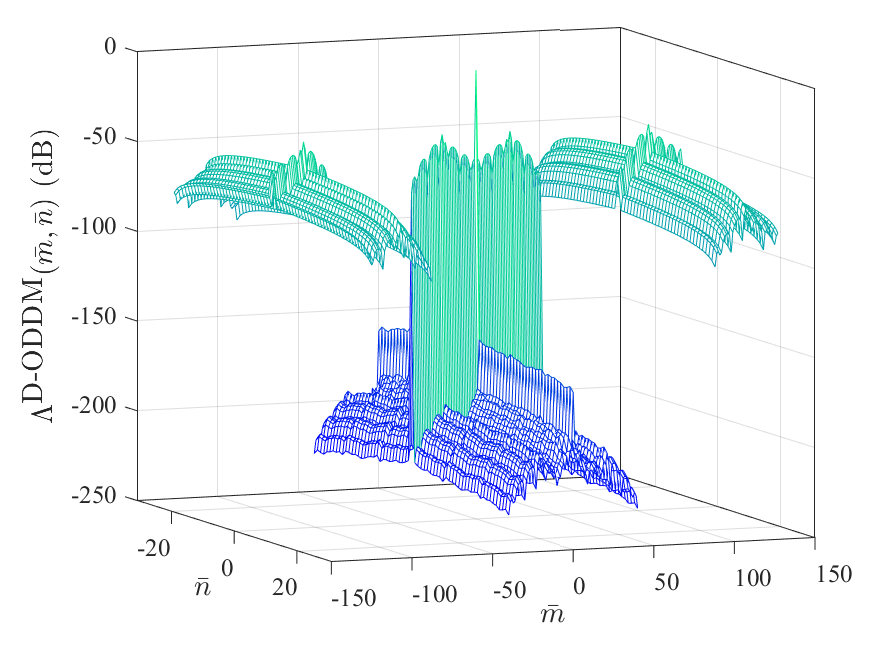}} \hspace{4 mm} 
  \subfloat[$\Lambda^{\textrm{D-ODDM}}(\bar{m},\bar{n})$ of {approximate digital} ODDM when $T_a=10T$.]{ \includegraphics[width=0.9\columnwidth]{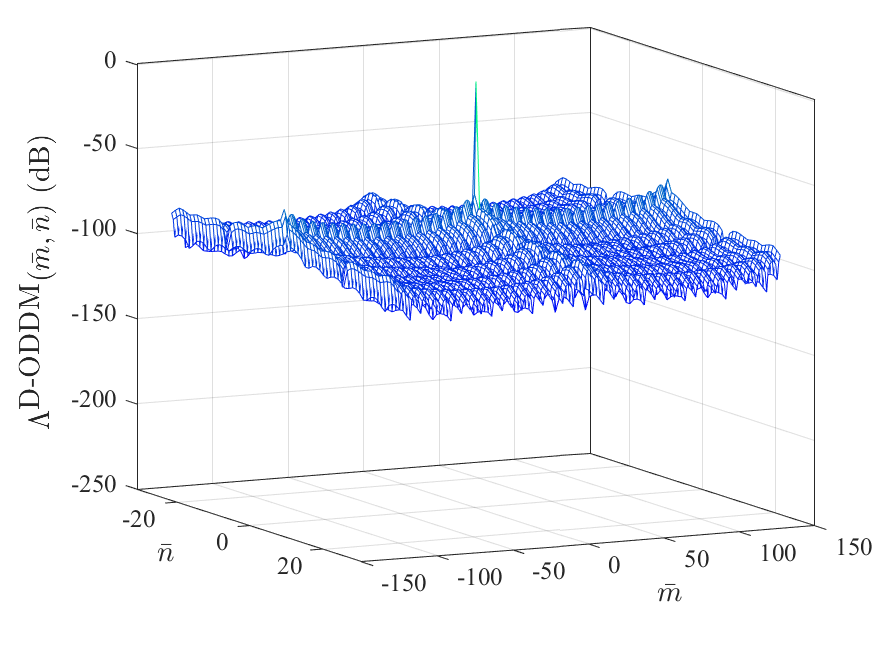}}
  \caption{Illustration of the orthogonality characteristics of {approximate digital} ODDM systems ($M=128$, $N=32$)}\label{Fig:NumOrthAmbuity-D-ODDM}
\end{figure}


 
Although the previous studies~\cite{2022_TWC_JH_ODDM,2023_ODDM_TCOM} have investigated the orthogonality characteristics of analog ODDM waveforms, to the best of the authors' knowledge, no studies have explored the orthogonality characteristics of {approximate digital} ODDM waveforms. To address this gap, we first derive the following theorem, which offers insights into the orthogonality characteristics of {approximate digital} ODDM waveforms.

\begin{theorem}\label{Thr:Orthonorm_D-ODDM}
For any SRN sub-pulse with duration $T_a\lesseqgtr T$, the basis functions of {approximate digital} ODDM waveforms form the orthonormal set of
\begin{align}\label{Equ:Orthonorm_D-ODDM}
\Psi^{\textrm{D-ODDM}}(m,m',n,n')
&=\delta (m'-m)\delta (n'-n), \notag\\&~~\forall m,m'\in\mathcal{S}_{0,M}, n',n\in\mathcal{S}_{0,N}. 
\end{align} 

\textit{Proof:} The proof is given in Appendix C.
\hfill $\blacksquare$
\end{theorem}

It has to be highlighted that Theorem \ref{Thr:Orthonorm_D-ODDM} for {approximate digital} ODDM waveforms has an important difference as compared to that in \eqref{Equ:Orthonorm_A-ODDM} for analog ODDM waveforms. 
In particular, 
while the result for analog ODDM waveforms in \eqref{Equ:Orthonorm_A-ODDM} was limited to the scenario where  $T_a\ll T$, the result in Theorem \ref{Thr:Orthonorm_D-ODDM} for {approximate digital} ODDM waveforms do not involve any such limitations on $T_a$. 
Due to this, it becomes possible to increase $T_a$ in {approximate digital} ODDM  waveforms without introducing {CS and an additional CP} as in analog ODDM waveforms. Considering this, \textit{we propose {approximate digital} ODDM waveforms without {CS and an additional CP} for all $T_a\lesseqgtr T$.}

We also highlight that Theorem \ref{Thr:Orthonorm_D-ODDM} for {approximate digital} ODDM waveforms involves the approximation \textit{Approx.~1}, but not \textit{Approx.~2}. 
However, similar to that in  \eqref{Equ:Orthonorm_A-ODDM}, the resulting impact of \textit{Approx.~1} on the validity of Theorem \ref{Thr:Orthonorm_D-ODDM} is negligible.  

Finally, we emphasize that despite the differences in the approximations associated with the derivations in \eqref{Equ:Orthonorm_A-ODDM} and Theorem \ref{Thr:Orthonorm_D-ODDM}, both analog and {approximate digital} ODDM waveforms satisfy the same orthonomal set given in \eqref{Equ:Orthonorm_A-ODDM} and Theorem \ref{Thr:Orthonorm_D-ODDM}. Consequently, \textit{analog and {approximate digital} ODDM systems share approximately the same orthogonality characteristics and the I/O relation when the physical channel taps have on-grid delays and Dopplers}. However, when the physical channel taps have off-grid delays and Dopplers, the analog and {approximate digital} ODDM systems may exhibit different I/O relations due to the differences in the way their transceivers interact with the channel, which needs to be investigated in future works~\cite{2024_TCOM_JunTong_ODDMoverPhysicalChannels,CAS_2025_C_ICC_Jun_ODDMinHighSpreadChannels}.



To understand the validity of Theorem 2, we next numerically evaluate  $\Lambda^{\textrm{D-ODDM}}(\bar{m},\bar{n})=\Lambda^{\textrm{D-ODDM}}(m'-m,n'-n)=\mathbb{E}_{\forall m,m'\in\mathcal{S}_{0,M}, n',n\in\mathcal{S}_{0,N}}[|\Psi^{\textrm{D-ODDM}}(m,m',n,n')|]$ and plot the results in Fig.~\ref{Fig:NumOrthAmbuity-D-ODDM}. Similar to that in Fig.~\ref{Fig:NumOrthAmbuity-A-ODDM}, we consider $T_a\approx 0.3T$ in Fig.~\ref{Fig:NumOrthAmbuity-D-ODDM}(a) and $T_a=10T$ in Fig.~\ref{Fig:NumOrthAmbuity-D-ODDM}(b). We 
first observe the validity of the orthogonality characteristics of {approximate digital} ODDM systems for any given $T_a\lesseqgtr T$.
Second, by comparing Fig.~\ref{Fig:NumOrthAmbuity-A-ODDM}(a) with Fig.~\ref{Fig:NumOrthAmbuity-D-ODDM}(a), we observe that when the sub-pulse duration is relatively small, the orthogonality characteristics of analog and {approximate digital} ODDM waveforms are similar. 
Finally, by comparing Fig.~\ref{Fig:NumOrthAmbuity-A-ODDM}(b) with Fig.~\ref{Fig:NumOrthAmbuity-D-ODDM}(b), we observe that when the sub-pulse duration is relatively large, the orthogonality characteristics of analog and {approximate digital} ODDM systems are similar,  except when $\bar{n} \neq 0$. For $\bar{n} \neq 0$, while $A_{u_{\textrm{ce}},u}(\bar{m}\mathcal{T},\bar{n}\mathcal{F})$ in Fig.~\ref{Fig:NumOrthAmbuity-A-ODDM}(b) was below $-150~\textrm{dB}$, the $\Lambda^{\textrm{D-ODDM}}(\bar{m},\bar{n})$ in Fig.~\ref{Fig:NumOrthAmbuity-D-ODDM}(b) is below $-80~\textrm{dB}$. This slightly better orthogonality characteristics of analog ODDM waveforms can be attributed to analog ODDM systems possessing {CS and an additional CP}, which {approximate digital} ODDM waveforms do not possess.
\section{Digital ODDM System and Other Variants of DD Modulation}



In this section, we compare the implementation of ODDM with other variants of DD modulation.
{We first briefly review the original OTFS proposal, its pulses, and its low-complex digital implementation~\cite{2017_WCNC_OTFS_Haddani,2018_TWC_Viterbo_OTFS_InterferenceCancellation}.} Thereafter, we provide a detailed examination of the similarities and differences that the transmitter of {{approximate digital} ODDM} systems share with those of other variants of DD modulation. Finally, we provide a brief comparison of the spectrum of DD
modulation variants.

\color{black}
\subsection{Orthogonal TF Space (OTFS) Modulation}

DD modulation was first introduced in the form of OTFS~\cite{2017_WCNC_OTFS_Haddani}. 
To enable effective DD modulation over doubly-selective LTV channels, it is crucial to achieve bi-orthogonality w.r.t. delay resolution $\mathcal{T}=T/M$ and Doppler resolution $\mathcal{F}=1/NT$, as it ensures minimal ISI in the DD domain, hence enabling simpler channel estimation and equalization~\cite{2023_ODDM_TCOM}. 

At the time of the original OTFS proposal, it was believed that pulses that could attain such bi-orthogonality would violate the Heisenberg uncertainty principle, and thus could not exist. To circumvent this, OTFS first introduced a 2D orthogonal precoding from the DD domain to the TF domain—the  inverse simplectic finite Fourier transform (ISFFT) precoder 
\begin{align}\label{xt_OTFS1} 
X_{\textrm{tf}}[\dot{n},\dot{m}] &= \frac{1}{\sqrt{NM}}\sum_{n=0}^{N-1} \sum_{m=0}^{M-1} X_{\textrm{dD}}[m,n]e^{j2\pi\left(\frac{n\dot{n}}{N}-\frac{m\dot{m}}{M}\right)},\notag\\ &~~~~~~~~~~~~~~~~~~~~~~~~~~\forall \dot{n}\in\mathcal{S}_{0,N},\dot{m}\in\mathcal{S}_{0,M}, \tag{28}
\end{align}
Thereafter, an MC modulation on the conventional TF plane (TFMC) (resolutions of $T$ in time and $1/T$ in frequency) was proposed to yield the OTFS transmit waveform as
\begin{align}\label{xt_OTFS2} \tag{29}
x(t)&=\sum _{\dot{n}=0}^{N-1}\sum _{\dot{m}=0}^{M-1}X_{\textrm{tf}}[\dot{n},\dot{m}]e^{j2\pi \frac{\dot{m}}{T} (t-\dot{n}T)}g_{\textrm{tf}}(t-\dot{n}T),
\end{align}
where  $g_{\textrm{tf}}(t)$ is the transmit pulse of OTFS.


\subsubsection{OTFS Transmit Pulse}

The ideal pulse,  which 
causes neither inter-carrier-interference (ICI) nor ISI, was proposed as the transmit pulse in the originally proposed  OTFS~\cite{2017_WCNC_OTFS_Haddani}. However, such ideal pulses are yet to be discovered. Given this, the $T$-length rectangular pulse  was first introduced in \cite{2018_TWC_Viterbo_OTFS_InterferenceCancellation} as the  transmit pulse for OTFS, i.e., $g_{\textrm{tf}}(t)=\Pi_T(t-\frac{T}{2})$,  and it has been widely adopted in subsequent OTFS studies~\cite{2018_TWC_Viterbo_OTFS_InterferenceCancellation,2020_TVT_Viterbo_RakeDFEforOTFS,2022_TWC_Lorenzo_OTFSvsOFDMcomparison,2021_WCM_JH_OTFS,2019_TCT_ReducedCPOTFS}.\footnote{
{
The key novelty of ODDM compared to the original OTFS framework lies in its basis function design. Specifically, the basis functions of ODDM are constructed to align directly with the inherent operations of doubly selective channels, whereas those of OTFS do not mirror such channel operations. Owing to this alignment of the basis functions of ODDM and the orthogonality of the DDOP w.r.t. delay and Doppler resolutions, ODDM ensures that the same set of orthogonal basis functions can sparsely represent the received signal at the output of a single-tap on-grid DD channel. This unique coupling between DD-domain information symbols and the DD channel results in minimal ISI in the DD domain—zero ISI for single-path on-grid channels and only $P$-tap ISI for $P$-path on-grid channels, i.e., no more than what is inherently introduced by the channel. This, in turn, leads to simplified channel estimation and equalization. In contrast, since OTFS basis functions are not inherently matched to the operations of doubly selective channels, OTFS suffers from additional ISI when using rectangular pulses over such channels (see \cite[Fig. 11]{2024_TCOM_JunTong_ODDMoverPhysicalChannels}). Moreover, as shown in \cite[Fig. 8]{2022_TWC_JH_ODDM}, and will be further emphasized in Fig. \ref{Fig:EPSD_Num}, ODDM can also offer improved OOBE compared to the original OTFS with rectangular pulses/windows.
}}

\subsubsection{Digital Implementation of OTFS {with Rectangular Pulses}}

\begin{figure*}[t]
\centering
\includegraphics[width=2\columnwidth]{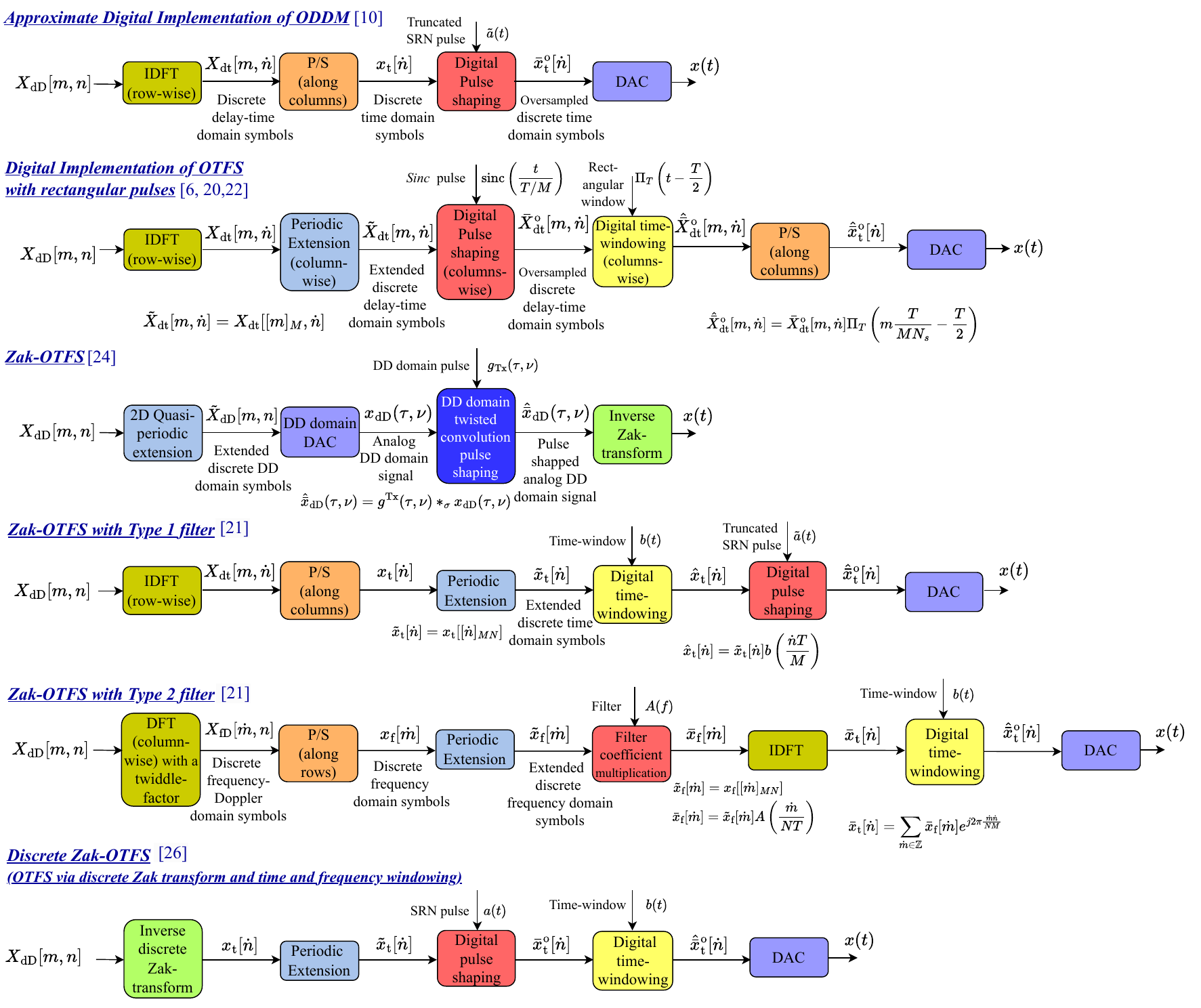}
\caption{Comparison of the transmitter implementation of ODDM with recently proposed other variants of DD modulation.  {For combating the channel delay spread, all the variants of DD modulation use the standard CP. However, for simplicity, this standard CP addition operations are avoided in this figure.}}\label{Fig:AllSignals} 
\end{figure*}

Similar to the challenges faced with analog ODDM, implementation of the aforementioned originally proposed OTFS with a rectangular pulse $\Pi_T(t - \frac{T}{2})$ is also difficult, as such an approach would necessitate multiple analog devices, including mixers in the MC modulator~\cite{2018_TWC_Viterbo_OTFS_InterferenceCancellation}.
However, such an analog implementation of OTFS can be simplified with low complexity by leveraging the inherent relationship between MC modulation/demodulation and the IDFT/DFT, without involving any additional approximation. 
In particular,  the $M$ mixers and adder operations in the TFMC modulator in \eqref{xt_OTFS2} can be simplified using an $M$-point IDFT, followed by cyclic extension and an ideal low-pass filter (LPF) with a passband bandwidth of $M/T$. Furthermore, the resulting $M$-point IDFT and the ISFFT in \eqref{xt_OTFS1}  can then be replaced with a single $N$-point IDFT.
These results in the DD modulation variant we refer to as digital implementation of OTFS, or simply digital OTFS in this work, and it is illustrated in the second subfigure of Fig. \ref{Fig:AllSignals}.

\color{black}

\subsection{Comparison of DD Modulation Variants}


We now investigate in detail the similarities and differences that the transmitters of  DD modulation variants share with each other. 
Apart from the {approximate digital} implementation of ODDM and the digital implementation of OTFS with rectangular pulse/window discussed above, we also focus on
(i) Zak-OTFS~\cite{2023_OTFS20_Paper1}, (ii) Zak-OTFS with Type 1 filter~\cite{2023_ZakOTFS_Implementation_Macquarie}, (iii) Zak-OTFS with Type 2 filter~\cite{2023_ZakOTFS_Implementation_Macquarie}, and (iv) OTFS via discrete-Zak transform and time and frequency windowing~\cite{2023_Shangyuan_ZakbasedPulseShapingforOTFS}, which we refer to as discrete-Zak OTFS. 
  A simplified schematic illustration of the transmitters of {{approximate digital} ODDM} and the aforementioned DD modulation variants is shown in Fig.~\ref{Fig:AllSignals}.\footnote{
{
In addition to the DD modulation variants detailed in Fig. \ref{Fig:AllSignals}, several other DD modulation variants have also been introduced in the literature. To the best of the authors' knowledge, these include the discrete Zak transform-based OTFS with rectangular transmitter and receiver window \cite{2022_EmanuelBook_DDCom}, circular pulse-shaped OTFS \cite{2023_Arman_PulseShapingforDDM}, linear pulse-shaped OTFS \cite{2023_Arman_PulseShapingforDDM}, and DD-OFDM~\cite{2023_OFDMbasedODDM_Macquarie_Final}. Note that these variants can be related to those discussed in Fig. \ref{Fig:AllSignals}. In particular, the version of linear pulse-shaped-OTFS \cite{2023_Arman_PulseShapingforDDM} in which only a single CP is prepended to its digital sequence shares similarities to the {approximate digital} implementation of ODDM.  Moreover, when greater flexibility is offered in the choice of the pulse shaping filter, the digital implementation of OTFS becomes the circular pulse-shaped OTFS \cite{2023_Arman_PulseShapingforDDM}. Furthermore, DD-OFDM can be viewed as the special case of Zak-OTFS with Type 2 filter when the considered time window is rectangular and does not incorporate a twiddle factor (i.e., the phase offset). As such, these DD modulation variants have not been included in Fig. \ref{Fig:AllSignals}, nor have they been discussed in detail in this section. 
}}

\begin{table*}[t]
\footnotesize 
\caption{Comparison of the transmitter implementations of different variants of DD modulation.}\vspace{-4.5mm}
\begin{center}
\begin{tabular}{|p{0.12\linewidth}|p{0.15\linewidth}|p{0.04\linewidth}|p{0.07\linewidth}|p{0.06\linewidth}|p{0.07\linewidth}|p{0.07\linewidth}|p{0.1\linewidth}|p{0.115\linewidth}|}
\hline \textbf{Approach}& \textbf{Domain Changes the} & \multicolumn{3}{|c|}{ \textbf{Pulse-shaping} } & \multicolumn{3}{|c|}{\textbf{Windowing}} & \textbf{Order of Pulse-} \\
\cline{3-8}
  &   \textbf{Symbols Undergo} & \textbf{Type} &  \textbf{Implement.  Domain} & \textbf{Function} & \textbf{Type} &  \textbf{Implement.  Domain} & \textbf{Function} & \textbf{shaping (PS) and windowing (W) } \\
\hline 
Digital Implem. of ODDM~\cite{2022_TWC_JH_ODDM} & \( X_{\textrm{dD}}[n,m] \) $\rightarrow$ Discrete time  $\rightarrow$ $x(t)$ & Direct &  Time &  $\tilde{a}(t)$ & Indirect & Time & Rectangular $\Pi_{NT}(t-\frac{NT}{2})$ &  W $\rightarrow$ PS \\
\hline
Digital Implem. of OTFS {with rect. pulse} \cite{2017_WCNC_OTFS_Haddani,2018_TWC_Viterbo_OTFS_InterferenceCancellation,2023_Arman_PulseShapingforDDM} & \( X_{\textrm{dD}}[n,m] \) $\rightarrow$ Discrete time  $\rightarrow$ $x(t)$ & Direct (Block-wise) &  Time &  \!~\!\!\!$\sinc(\frac{t}{T/M})$ & Direct (Multiple) & Time & Rectangular $\Pi_{T}(t-\frac{T}{2})$ &  PS $\rightarrow$ W \\
\hline
Zak-OTFS~\cite{2023_OTFS20_Paper1}  & \( X_{\textrm{dD}}[n,m] \) $\rightarrow$ Oversampled DD  $\rightarrow$ $x(t)$ & Indirect &  DD & $\tilde{a}(t)$ & Indirect & DD &  $b(t)$ &  -\\
\hline
Zak-OTFS with Type 1 filter~\cite{2023_ZakOTFS_Implementation_Macquarie} & \( X_{\textrm{dD}}[n,m] \) $\rightarrow$ Discrete time  $\rightarrow$ $x(t)$ & Direct &  Time &  $\tilde{a}(t)$ & Direct & Time & $b(t)$ &  W $\rightarrow$ PS \\
\hline
Zak-OTFS with Type 2 filter~\cite{2023_ZakOTFS_Implementation_Macquarie} & \( X_{\textrm{dD}}[n,m] \) $\rightarrow$ Discrete frequency  $\rightarrow$ $x(t)$ & Direct &  Frequency  &   $a(t)$ & Direct & Time & $b(t)$ & PS $\rightarrow$ W \\
\hline
Discrete-Zak-OTFS~\cite{2023_Shangyuan_ZakbasedPulseShapingforOTFS} & \( X_{\textrm{dD}}[n,m] \) $\rightarrow$ Discrete time  $\rightarrow$ $x(t)$ & Direct &  Time &  $a(t)$ & Direct & Time & $b(t)$ & PS $\rightarrow$ W \\
\hline
\end{tabular}\label{tab:DDMCompare}
\end{center}
\end{table*}


We observe that, like the {{approximate digital} ODDM} transmitter, the transmitters of 
all the DD modulation variants involve performing all their transformations,  pulse shaping, and/or windowing in the digital domain. However, as summarized in Table \ref{tab:DDMCompare}, the similarities and differences among them arise based on the domain changes the symbols undergo, the type of pulse shaping and windowing used, and the domains and sequence in which they are performed.


\textbf{Domain Changes the Symbols Undergo:}
In {{approximate digital} ODDM} and digital OTFS, Zak-OTFS with Type 1 filter, and discrete-Zak-OTFS, the DD domain symbols \( X_{\textrm{dD}}[n,m] \) are first converted to the discrete time domain, and then to the continuous time domain $x(t)$. 
Conversely,  in Zak-OTFS, the DD domain symbols \( X_{\textrm{dD}}[n,m] \) are first converted to the continuous-DD domain, and then to the continuous time domain. In Zak-OTFS with Type 2 filter, the DD domain symbols \( X_{\textrm{dD}}[n,m] \) are first converted to the discrete frequency domain,   then to the discrete time domain, and finally to the continuous time domain.

\textbf{Pulse-shaping and Windowing:}
The {{approximate digital} ODDM}, Zak-OTFS with Type 1 and 2 filters, and discrete-Zak OTFS employ $a(t)$-based pulse shaping to limit the transmit-ready {waveform} in the frequency domain. In contrast, the digital OTFS employs $\sinc \left(\frac{t}{T/M}\right)$-based pulse shaping, although digitally pulse-shaped OTFS can be considered to have $a(t)$-based pulse shaping~\cite{2023_Arman_PulseShapingforDDM}.



It has to be noted that the {{approximate digital} ODDM} does not explicitly use a window to limit the transmit {waveform} in time. However, given the fact that the discrete time-domain symbols $x_{\textrm{t}}[\dot{n}]$ on which pulse shaping is performed are finite in length, {{approximate digital} ODDM} 
indirectly employs a single rectangular window $\Pi_{NT}(t-\frac{NT}{2})$ with duration $NT$.
The digital OTFS directly employs multiple rectangular windows $\Pi_T(t-\frac{T}{2})$ of duration $T$ (i.e., block-wise windowing). 
As for Zak-OTFS with Type 1 and 2 filters, and discrete-Zak OTFS, they employ $b(t)$-based windowing.  The $b(t)$ can either be a rectangular window with the duration $NT$ or a non-rectangular window with the duration $NT(1+\beta_t)$, where $\beta_t$ defines the excess duration and the time-domain roll-off factor. 


The Zak-OTFS does not directly employ time-domain pulse shaping or windowing. However, it employs DD-domain twisted convolution and pulse shaping in the delay and Doppler domains corresponds to limiting the {waveform} in the frequency and time, respectively~\cite{2023_ZakOTFS_Implementation_Macquarie,2023_Shangyuan_ZakbasedPulseShapingforOTFS}. {Due to these factors,} the 
Zak-OTFS can be viewed as if it indirectly employs pulse shaping and windowing~\cite{2023_ZakOTFS_Implementation_Macquarie}.

In terms of the domain in which pulse shaping operation is performed, all the discussed DD modulation variants, except Zak-OTFS and Zak-OTFS with Type 2 filter, employ pulse shaping operation in the time domain. Differently, the Zak-OTFS with Type 2 filter employs pulse shaping in the frequency domain. As for the domain of windowing, all the discussed DD modulation variants, except Zak-OTFS, employ a windowing operation in the time domain. 

When focusing on the order in which pulse shaping and windowing are performed, the {{approximate digital} ODDM} and Zak-OTFS with Type 1 filter employ windowing first, followed by pulse shaping. Differently, digital OTFS, Zak-OTFS with Type 2 filter, and discrete-Zak OTFS employ pulse shaping first, followed by windowing.\footnote{When pulse shaping is performed after windowing, as in the {{approximate digital} ODDM} and Zak-OTFS with Type 1 filter, the transmit waveforms will not be strictly time-limited if the SRN pulse used for pulse shaping is not strictly limited in time. To counter this, ODDM and Zak-OTFS with Type 1 filter employ pulse shaping based on the truncated SRN pulse~\cite{2023_ZakOTFS_Implementation_Macquarie,2022_TWC_JH_ODDM}.}

Finally, based on the above comparison, we can infer the following: 
The Zak-OTFS with Type 1 filter is the generalization of {approximate digital} ODDM to the scenario where the window is not necessarily a rectangular window.\footnote{{Given the similarities among {approximate digital} ODDM~\cite{2022_TWC_JH_ODDM}, Zak-OTFS with Type 1 filter~\cite{2023_ZakOTFS_Implementation_Macquarie}, and linear pulse-shaped OTFS~\cite{2023_Arman_PulseShapingforDDM}, it may be possible to unify these three DD modulation variants under a single framework. Moreover, the study~\cite{2023_Arman_PulseShapingforDDM} has already unified digital OTFS~\cite{2017_WCNC_OTFS_Haddani}, {approximate digital} ODDM~\cite{2022_TWC_JH_ODDM}, and linear pulse-shaped OTFS under one framework. Despite these, further investigation may be necessary to generalize/unify all DD modulation variants under a single global unified framework, which will be considered in our future works.}} 
Compared with the {{approximate digital} ODDM} and Zak-OTFS with Type 1 filter, the Zak-OTFS with Type 2 filter and discrete-Zak OTFS change the order in which pulse shaping and windowing are performed. Compared with the {{approximate digital} ODDM}, Zak-OTFS with Type 1 filter, and discrete-Zak OTFS, the Zak-OTFS with Type 2 filter changes the domain in which pulse shaping operation is performed.
Finally, when different DD domain twisted convolution filters are considered, Zak-OTFS serves as the DD domain representation of {approximate digital} ODDM, Zak-OTFS with Type 1 and 2 filters, and discrete-Zak OTFS.\footnote{\color{black}
Chirp-based MC modulations, such as AFDM~\cite{2023_TWC_AliBemani_AFDM} and OCDM~\cite{2016_TCOM_OCDM}, are formulated and evaluated using digital sequences. As such, they share similarities with the {approximate digital} implementation of ODDM, but differ in the transforms used to map symbols from the modulation domain to the time domain.  For example, at their transmitters, AFDM and OCDM use the inverse discrete affine Fourier transform (IDAFT) and inverse discrete Fresnel transform (IDFnT), respectively. 
Detailed comparisons of the implementation of DD modulation with chirp/affine-based designs will be considered in our future works. 
\color{black}
}


\color{black}
\subsection{Spectrum of Other DD Modulation Variants}

\color{black}
Due to the similarities and differences between {approximate digital} ODDM and other DD modulation variants, their spectrum also show both commonalities and distinctions. 
In particrular, similar to {{approximate digital} ODDM}, the spectrum of other DD modulation variants is confined to the power spectrum of their respective sub-pulses, though with varying levels of OOBE. As shown in~\cite{2022_TWC_JH_ODDM,2023_ZakOTFS_Implementation_Macquarie,2023_Shangyuan_ZakbasedPulseShapingforOTFS,2023_OFDMbasedODDM_Macquarie_Final}, OTFS exhibits higher OOBE levels compared to all the other DD modulation variants. When system parameters are kept consistent, Zak-OTFS with Type 1 and Type 2 filters, as well as Discrete-Zak-OTFS, demonstrate similar OOBE compared to ODDM~\cite{2023_ZakOTFS_Implementation_Macquarie,2023_Shangyuan_ZakbasedPulseShapingforOTFS,2023_OFDMbasedODDM_Macquarie_Final}. 

\color{black}


\section{Simulation Results}
\label{Sec:Num}

In this section, we present numerical results to compare the spectrum and BER of analog and {approximate digital} ODDM systems. 
Unless specified otherwise, we consider $M=128$, $N=32$, 4-QAM waveforming, $T\triangleq\frac{1}{\Delta f}=\frac{1}{15~\mathrm{kHz}}$, the SRRC  pulse as the SRN sub-pulse with $\beta=0.15$, $T_a\approx 0.3T$, {and $T_{\textrm{cp}}=\frac{T}{10}=6.67 \mu \textrm{s}>\tau_{\mathrm{max}}=2.51 \mu \textrm{s}$}~\cite{2023_ODDM_TCOM}.

\begin{figure*}[t]
\centering
\includegraphics[width=2\columnwidth]{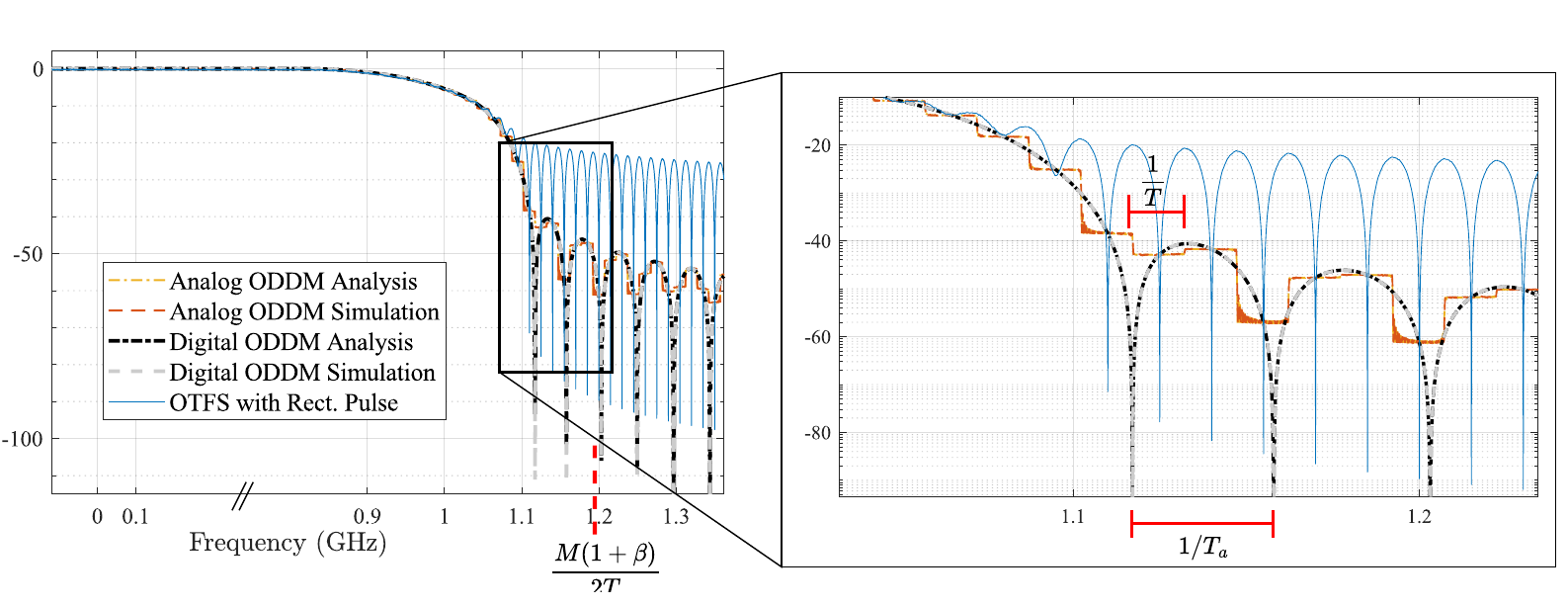}
\caption{Illustration of the one-sided power spectral densities of the waveforms in analog ODDM systems,  {approximate digital} ODDM systems, and OTFS with rectangular pulses $\Pi(t-\frac{T}{2})$.}\label{Fig:EPSD_Num} 
\end{figure*}

In Fig.~\ref{Fig:EPSD_Num}, we numerically evaluate the {spectra} of analog and {approximate digital} ODDM waveforms.
To this end, we plot the analytical expected power spectrum densities derived in \emph{Theorem \ref{Thr:EPS_ODDM}} and the numerically calculated expected power spectrum densities of the simulated analog and {approximate digital} ODDM waveforms. For comparison, the expected power spectrum density of OTFS waveforms {with RRC-based pulse-shaping}~\cite{2017_WCNC_OTFS_Haddani} is also plotted.
We first observe that the analytical and numerical results match well, thereby validating our derivations in \emph{Theorem \ref{Thr:EPS_ODDM}}.
Moreover, we observe that, as expected,  the spectrum of analog ODDM waveforms exhibits a step-wise behavior in its transition region with a step width of $1/T$. Furthermore, we observe that the side-lobes in the frequency domain of {approximate digital} ODDM waveforms have a width of $\frac{1}{T_a}$), since its spectrum is dictated by $|\tilde{A}(f)|^2$ and $\tilde{A}(f)$ has side-lobes due to the truncation of the sub-pulse in the time domain. Additionally, we observe that the side-lobes of waveforms in both analog and {approximate digital} ODDM waveforms are significantly lower than that of OTFS, demonstrating the superior OOBE performance for ODDM.

\begin{figure}[t]
\centering
\includegraphics[width=0.99\columnwidth]{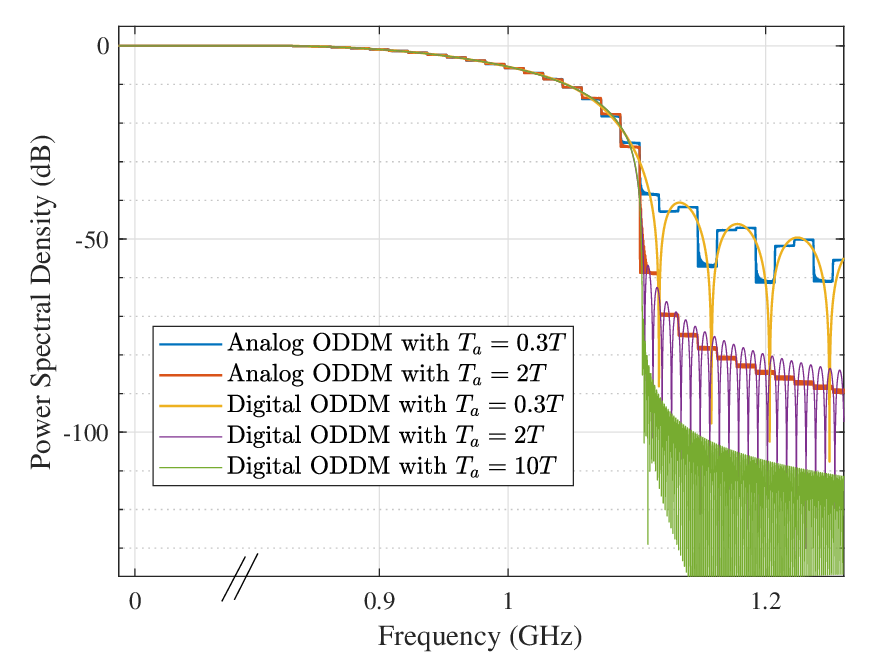}
\caption{Illustration of the impact of increasing $T_a$ on the power spectral densities of the waveforms in analog and {approximate digital} ODDM systems.}\label{Fig:EPSD_Num2} 
\end{figure}

\begin{figure}[t]
\centering
\includegraphics[width=0.99\columnwidth]{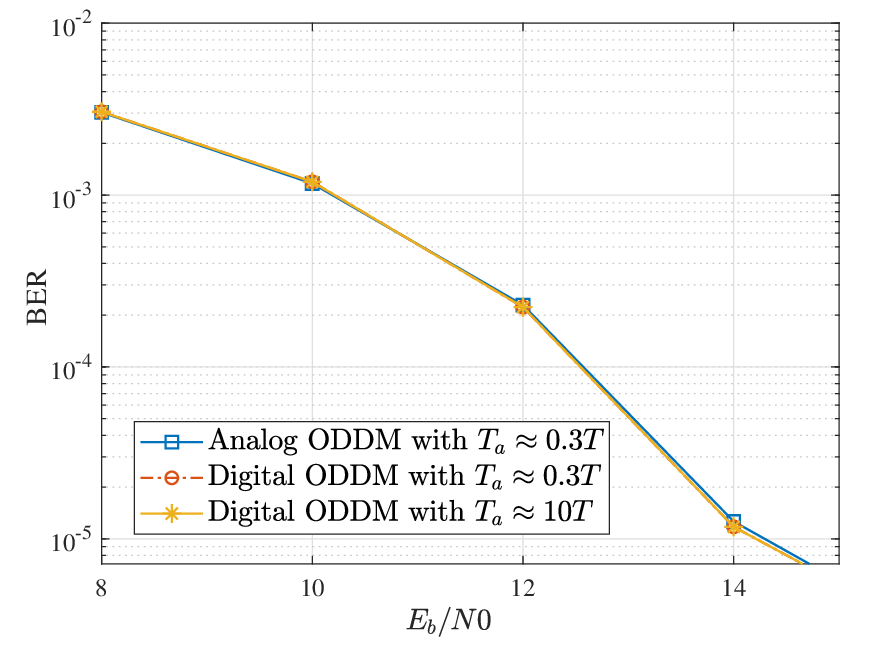}
\caption{Illustration of the BER versus $E_{b}/N_0$.}\label{Fig:NumBER} 
\end{figure}


When examining the $\Upsilon$-dB bandwidth of both analog and {approximate digital} ODDM waveforms in Fig.~\ref{Fig:EPSD_Num} for all $\Upsilon \in \{3, 7, 10, 20, 30, 40\}$, we found that $\Upsilon$-dB bandwidth of both are almost the same, with the maximum difference being upper bounded by \(\frac{1}{2T}\). Besides, we observe that despite the difference in the behavior of their sidelobes, the maximum of the sidelobes and the effective energy under all the sidelobes in the OOBE region are almost the same for both. This shows that \textit{the effective spectral occupancy of analog and {approximate digital} ODDM waveforms are the same.} 


In Fig. \ref{Fig:EPSD_Num2}, we plot the expected power spectrum densities of analog and {approximate digital} ODDM waveforms for different values of $T_a$. We observe that as $T_a$ increases, the sidelobes of both analog and {approximate digital} ODDM waveforms roll off rapidly, leading to improvements in their OOBE performance. 
This enhancement in OOBE performance with increasing $T_a$  illustrates the trade-off between the occupied time and frequency domain resources of ODDM waveforms.

Finally, in Fig.~\ref{Fig:NumBER}, we examine the bit error rate (BER) performance of analog and {approximate digital} ODDM systems versus $E_{b}/N_0$. 
We let $T_{\textrm{cp}}=\frac{T}{8}$, and use the channel EVA model while considering the carrier frequency of 5 GHz and the user speed of 500 km/h~\cite{2018_TWC_Viterbo_OTFS_InterferenceCancellation}. At the receiver, we employ the MP algorithm for {signal equalization and detection}~\cite{2018_TWC_Viterbo_OTFS_InterferenceCancellation}. 
First, we observe that when $T_a$ is low, the BER of analog ODDM system is the same as that of {approximate digital} ODDM systems for all $E_{b}/N_0$ values. This can be attributed to the similar orthogonality characteristics possessed by both analog and {approximate digital} ODDM systems when $T_a \ll T$, as highlighted in Figs.~\ref{Fig:NumOrthAmbuity-A-ODDM}(a) and~\ref{Fig:NumOrthAmbuity-D-ODDM}(a). 
Second, by comparing the plots for {approximate digital} ODDM systems for varying $T_a$, we see that the BER performance of {approximate digital} ODDM systems remains the same despite the increase in $T_a$. This observation indicates that although increasing $T_a$ slightly improves the orthogonality characteristics of {approximate digital} ODDM systems at the expense of longer transmit {waveform} duration (see Figs.~\ref{Fig:NumOrthAmbuity-A-ODDM} and~\ref{Fig:NumOrthAmbuity-D-ODDM}), the resulting impact on the BER of {approximate digital} ODDM systems is negligible.
Based on this result and that in Fig. \ref{Fig:EPSD_Num2}, we conclude that the exact $T_a$ for the ODDM sub-pulse should be determined by considering the tradeoff between duration and OOBE of transmit waveforms, while disregarding changes in orthogonality.

\section{Conclusion}

In this work, we investigated analog and {approximate digital} ODDM systems, in terms of their implementation, time and frequency domain representations of their transmit waveforms, orthogonality characteristics, and the similarities and differences that they share with the systems of other DD modulation variants.  The main conclusions drawn from our investigation are as follows: 

\begin{itemize}
      \item  {The spectrum of analog ODDM waveforms exhibits a step-wise behavior in its transition regions, whereas the spectrum of {approximate digital} ODDM waveforms is confined to that of the ODDM sub-pulse.}
  \item {Unlike analog ODDM waveforms, the {approximate digital} ODDM  waveforms can attain orthogonality w.r.t. delay and Doppler resolutions without necessitating {additional time domain cyclic extensions,} 
      even when the sub-pulse duration is high.}
  \item Similarities and differences between the {approximate digital} ODDM system and of other variants of DD modulation stem from the domain changes the symbols undergo, the type of pulse shaping and windowing used, and the domain and sequence in which they are applied.
  \item The {approximate digital} ODDM system achieves similar BER performance compared to its analog counterpart while requiring significantly lower implementation complexity. 
\end{itemize}

\appendices

\color{black}
\section{Different Variants of Digital ODDM Transmitter} \label{App:DigitalODDM}

\begin{figure}[t]
\centering
\includegraphics[width=0.99\columnwidth]{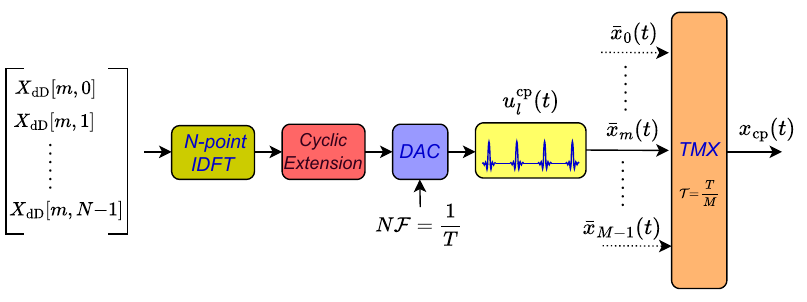}
\caption{{Illustrations of the direct digital implementation of ODDM transmitter with the DDOP $u(t)$ considered as prototype pulse/window function.}}\label{Fig:A-ODDM_1}
\end{figure}

Leveraging the connection between MC modulation and the IDFT \cite{1971_Weinstein_OFDM_usingDFT,1967_IEEETCT_zimmerman_OFDM_usingDFT}, the mixer and the adder operations in the analog ODDM transmitter in Fig. \ref{Fig:A-ODDM}(a) can be replaced with IDFT,  cyclic extension, and an ideal LPF with passband bandwidth ${N}\mathcal F=\frac{1}{T}$ \cite{2022_ICC_JH_ODDM}, which is a DAC having a rate of ${N}\mathcal F=\frac{1}{T}$. The resulting direct digital implementation of the ODDM transmitter is shown in Fig. \ref{Fig:A-ODDM_1}. 
This direct digital ODDM transmitter achieves lower complexity compared to the analog ODDM transmitter shown in Fig. \ref{Fig:A-ODDM}(a). However, it still entails notable complexity due to the need for truncation/windowing and time multiplexing based on DDOP $u(t)$ which has a relatively longer duration.

\begin{figure}[t]
\centering
\includegraphics[width=0.99\columnwidth]{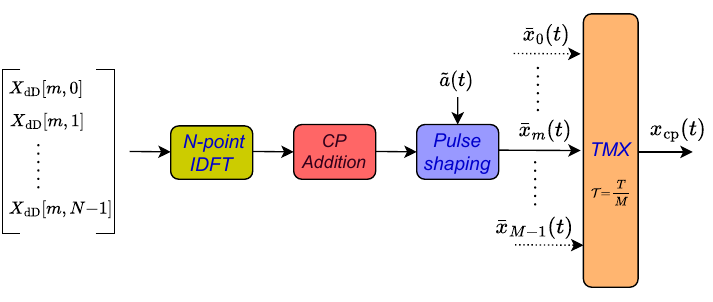}
\caption{{Illustrations of the {approximate digital} implementation of ODDM  transmitter with multiple branches of $a(t)$-based filtering/pulse-shaping.}}\label{Fig:A-ODDM_2}
\end{figure}

Recall that the DDOP $u(t)$ is a collection of multiple sub-pulses $a(t)$ that has a limited duration of $T_a=2Q\frac{T}{M}\ll T$. As a result,  waveforms truncated/windowed based on $u(t)$
become discontinuous with $N$ segments, each with a length $T_a$. This enables the DAC with the rate of ${N}\mathcal F=\frac{1}{T}$ and the $u(t)$-based truncation/windowing operations to be \emph{approximated} with $a(t)$-based filtering/sample-wise pulse-shaping. 
This 
leads to the {approximate digital} implementation of the ODDM transmitter shown in Fig. \ref{Fig:A-ODDM_2}. 

Note that the $M$ branches in Fig. \ref{Fig:A-ODDM_2} share the same $a(t)$-based filter. Thus, the order of the TMX and the filters can be exchanged to further simplify the implementation. 
Furthermore,  
the $a(t)$-based filter in Fig. \ref{Fig:A-ODDM_2} can also be implemented digitally, followed by a DAC having a high enough sampling rate, thus overcoming the need for analog circuits to implement the filter. These modifications result in the simplified low-complex {approximate digital} implementation of the ODDM transmitter shown in Fig. \ref{Fig:D-ODDM}(a).

\color{black}

\section{Proof of Lemma~\ref{Lemma:Phi_ODDM}}\label{App:Phi_ODDM}

By applying the time-shifting and frequency-shifting properties of Fourier transform on $\phi_{m,n}^{\textrm{A-ODDM}}(t)$ in \eqref{Equ:phi_A-ODDM}, $\Phi_{m,n}^{\textrm{A-ODDM}}(f)$ can be obtained as $e^{-j2\pi  fm\frac{T}{M}}U\left(f-\frac{\psi_N(n)}{NT}\right)$. 
Thereafter, by substituting $U(f)$ in \eqref{Equ:Uf} {into the above expression,}  $\Phi_{m,n}^{\textrm{A-ODDM}}(f)$ can be simplified as in (\ref{Equ:Phi_A-ODDM}).

To derive $\Phi_{m,n}^{\textrm{D-ODDM}}(f)$, we represent $\phi_{m,n}^{\textrm{D-ODDM}}(t)$ in \eqref{Equ:phi_D-ODDM} as 
\begin{align} \label{Equ:Phi_D-ODDM_proof1}
\phi_{m,n}^{\textrm{D-ODDM}}(t)=\frac{1}{\sqrt{N}}\tilde{a}(t) \star b_1(t-m\frac{T}{M}),
\end{align} 
where $\star $ denotes the linear convolution operation and $b_1(t)\triangleq e^{j2\pi  \frac{\psi_N(n)}{NT}t}b_2(t)b_3(t)$ with $b_2(t)\triangleq \sum_{\grave{n}=-\infty}^{\infty}\delta(t-\grave{n}T)$ and $b_3(t)\triangleq \Pi_{NT}\left(t-\frac{(N-1)T}{2}\right)$.  Thereafter, by applying the time-shifting, convolution, and multiplication properties of Fourier transform on \eqref{Equ:Phi_D-ODDM_proof1}, $\Phi_{m,n}^{\textrm{D-ODDM}}(f)$ can be obtained as 
\begin{align} \label{Equ:Phi_D-ODDM_proof2}
&\Phi_{m,n}^{\textrm{D-ODDM}}(f)
=\tilde{A}(f)   B_1(f) e^{-j2\pi  fm\frac{T}{M}}\notag\\
&=\tilde{A}(f) \times  \left(B_2\left(f{-}\frac{\psi_N(n)}{NT}\right)\star B_3(f)\right) e^{-j2\pi  fm\frac{T}{M}},
\end{align} 
\begin{figure*}[b]
\normalsize 
\hrulefill
\begin{align} \label{Eq:EPSD_ODDM_proof0s}
&\mathbb{E}\left[\left|X^{\Psi\textrm{-ODDM}}(f)\right|^2\right]
=E_s\sum_{m=0}^{M-1}\sum_{n=0}^{N-1} \Phi_{m,n}^{\Psi\textrm{-ODDM}}(f)\Phi_{m,n}^{\Psi\textrm{-ODDM}}(f)^*.\\
&\mathbb{E}\left[\left|X^{\textrm{A-ODDM}}(f)\right|^2\right]\notag\\ 
&=E_sNM\sum_{\tilde{n}=-\frac{N}{2}}^{\frac{N}{2}-1} \left| \tilde{A}(f-\frac{\tilde{n}}{NT}) \right|^2 \sum_{\dot{m}=-\infty}^{\infty}\sum_{\grave{m}=-\infty}^{\infty}e^{j \pi (N-1)(\dot{m}-\grave{m})}\sinc(NTf-\dot{m}N-\tilde{n})\sinc(NTf-\grave{m}N-\tilde{n}).\label{Eq:EPSD_ODDM_proof4}\\
&\mathbb{E}\left[\left|X^{\textrm{D-ODDM}}(f)\right|^2\right] 
=E_sNM \left| \tilde{A}(f) \right|^2\sum_{\tilde{n}=-\frac{N}{2}}^{\frac{N}{2}-1}  \sum_{\grave{m}=-\infty}^{\infty}\sinc^2(NTf-\grave{m}N-\tilde{n}).\label{Eq:EPSD_ODDM_proof7}
\end{align} 
\end{figure*}

\noindent
where $B_{\iota}(f)$ is the {Fourier/frequency domain representation} of $b_{\iota}(t)$ with $\iota\in\{1,2,3\}$. 
The $B_2(f)$ and $B_3(f)$ can be derived as $B_2(f)=\sum_{\grave{n}={-}\infty}^{\infty}e^{{-}j2\pi \grave{n}T f}{=}\frac{1}{T}\sum_{\dot{m}={-}\infty}^{\infty}\delta(f{-}\frac{\dot{m}}{T})$~and $B_3(f)=NTe^{{-}j\pi (N{-}1)Tf}\sinc(NTf)$. By substituting $B_2(f)$ and $B_3(f)$ in \eqref{Equ:Phi_D-ODDM_proof2} and then simplifying,  $\Phi_{m,n}^{\textrm{D-ODDM}}(f)$ can be obtained as in \eqref{Equ:Phi_D-ODDM}, which completes the proof. 
\hfill $\blacksquare$

\section{Proof of Theorem~\ref{Thr:EPS_ODDM}}\label{App:EPS_ODDM}


We start by expanding the expression for $\mathbb{E}\left[\left|X^{\Psi\textrm{-ODDM}}(f)\right|^2\right]$ using \eqref{Equ:Xf_ODDM} and $E[X_{\textrm{dD}}[m,n]X^*_{\textrm{dD}}[m',n']] = E_s\delta(m'-m)\delta(n'-n)$ to obtain \eqref{Eq:EPSD_ODDM_proof0s}. The expressions \eqref{Eq:EPSD_ODDM_proof0s}-\eqref{Eq:EPSD_ODDM_proof7} are given at the end of this page.

To obtain $\mathbb{E}\left[\left|X^{\textrm{A-ODDM}}(f)\right|^2\right]$, we first substitute $\Phi_{m,n}^{\textrm{A-ODDM}}(f)$ derived  in Lemma \ref{Lemma:Phi_ODDM}, into \eqref{Eq:EPSD_ODDM_proof0s}. In doing so, we obtain 
\eqref{Eq:EPSD_ODDM_proof4}. 
Next, we observe that 
the energy of the sinc tone $\sinc(NTf-mN-\tilde{n})$ that overlap with that of $\sinc(NTf-\grave{m}N-\tilde{n})$ is ignorable when $\dot{m}\neq \grave{m}$. Considering this, \eqref{Eq:EPSD_ODDM_proof4} is simplified as \eqref{Eq:EPSD_A-ODDM}.

To obtain $\mathbb{E}\left[\left|X^{\textrm{D-ODDM}}(f)\right|^2\right]$, we follow steps similar to those used to obtain \eqref{Eq:EPSD_A-ODDM} from \eqref{Eq:EPSD_ODDM_proof0s} while using $\Phi_{m,n}^{\textrm{D-ODDM}}(f)$ derived in Lemma \ref{Lemma:Phi_ODDM}. In doing so, we obtain \eqref{Eq:EPSD_ODDM_proof7}. Thereafter by rearranging \eqref{Eq:EPSD_ODDM_proof7} while considering $\dot{m}=\grave{m}N+\tilde{n}$, we arrive at \eqref{Eq:EPSD_D-ODDM}, which completes the proof. 
\hfill $\blacksquare$

\section{Proof of Theorem~\ref{Thr:Orthonorm_D-ODDM}}\label{App:Orthonorm_D-ODDM}

The goal is to derive \eqref{Equ:Orthonorm_D-ODDM} for all possible positive values of $T_a$. With this in mind, 
we first derive \eqref{Equ:Orthonorm_D-ODDM} for a special case of  $T_a$ (i.e., \textit{SC}). 
Based on this, we then derive \eqref{Equ:Orthonorm_D-ODDM} for two general cases of $T_a$ {(i.e., \textit{GC-I} and \textit{GC-II})}, which captures all possibilities of $T_a$.

Using $\phi_{m,n}^{\textrm{D-ODDM}}(t)$ in \eqref{Equ:phi_D-ODDM}, we first obtain 
\begin{align}\label{Equ:Orthonorm_D-ODDM_proof1}
&\Psi^{\textrm{D-ODDM}}(m,m',n,n')
=\int \phi_{m,n}^{\textrm{A-ODDM}}(t) \phi_{m',n'}^{\textrm{A-ODDM}}(t) \textrm{d}t \notag\\
&~~~~~~~~~=\frac{1}{N}\sum_{\dot{n}=0}^{N{-}1} \sum_{\grave{n}=0}^{N{-}1}e^{{-}j2\pi \frac{(n'\grave{n}{-}n\dot{n})}{N}}\notag\\
&~~~~~~~~~~~~~~~~\times \int\tilde{a}\left(\bar{t}{-}\dot{n}T\right) \tilde{a}^*\left(\bar{t}{-}\grave{n}T{-}\frac{\bar{m}T}{M}\right) \textrm{d}\bar{t}, 
\end{align} 
where $\bar{m}=m'-m$ and $\forall \bar{m}\in\mathcal{S}_{M-1,M}$.
Let $T_a$ be $T_a\triangleq T_{a1}+T_{a2}$, where $T_{a1}=KT$ with $K$ being a non-negative integer and $0\leq T_{a2}<T$. Consider the special case and the general cases as follows: In \textit{SC},  $T_{a1}=KT$ with $K> 0$ and $T_{a2}=0$, thus $T_a=KT\geq T$. In \textit{GC-I}, $T_{a1}=KT$ with $K\geq 0$ and $T_{a2}\leqslant \frac{T}{2}$. 
In \textit{GC-II}, $T_{a1}=KT$ with $K\geq 0$ and $T_{a2}>\frac{T}{2}$. 

Now we focus on \textit{SC} with $m\geqslant 0$. 
It can be shown that $\tilde{a}\left(\bar{t}-\dot{n}T\right)$ and $\tilde{a}\left(\bar{t}-\grave{n}T-\frac{\bar{m}T}{M}\right)$ in \eqref{Equ:Orthonorm_D-ODDM_proof1} overlap only when $\grave{n}=\max\{0,\dot{n}-K\}, \cdots, \min\{\dot{n}+K{-}1,N-1\}$ for all $\grave{n}\in\mathcal{S}_{0,N}$. Using this in \eqref{Equ:Orthonorm_D-ODDM_proof1} and rearranging it, 
we obtain 
\begin{align}\label{Equ:Orthonorm_D-ODDM_proof5}
\Psi^{\textrm{D-ODDM}}(m,m',n,n')=\Theta_0{+}\sum_{i=1}^K \Theta^+_{i}{+}\mathrm{nz}(K{-}1)\sum_{i=1}^{K{-}1} \Theta^-_{i}, 
\end{align} 

\noindent
where 
\begin{align}
\Theta_{0}&{=}\underbrace{\frac{1}{N}\sum_{\dot{n}=0}^{N-1} e^{-j2\pi \frac{\bar{n}\dot{n}}{N}} }_{\substack{\textrm{T}^0_{1}}} \underbrace{
\int_{-\frac{T_a}{2}}^{\frac{T_a}{2}}\tilde{a}\left(\bar{\bar{t}}\right) \tilde{a}^*\left(\bar{\bar{t}}-\frac{\bar{m}T}{M}\right) \textrm{d}\bar{\bar{t}}}_{\substack{\textrm{T}^0_{2}}},\notag\\
\Theta^{+}_{i}&{=}\frac{e^{j2\pi \frac{n'i}{N}}}{N} \sum_{\dot{n}=i}^{N-1} e^{-j2\pi \frac{\bar{n}\dot{n}}{N}}  \underbrace{\int\limits_{-\frac{T_a}{2}}^{\frac{T_a}{2}}\tilde{a}\left(\bar{\bar{t}}\right) \tilde{a}^*\left(\bar{\bar{t}}{-}iT{-}\frac{\bar{m}T}{M}\right)\textrm{d}\bar{\bar{t}}}_{\substack{\textrm{T}^{-}_{2}}},\notag\\
\Theta^{-}_{i}&{=}\frac{e^{-j2\pi \frac{n'i}{N}}}{N} \sum_{\dot{n}=0}^{N-1-i} e^{-j2\pi \frac{\bar{n}\dot{n}}{N}}  \underbrace{\int\limits_{-\frac{T_a}{2}}^{\frac{T_a}{2}}\tilde{a}\left(\bar{\bar{t}}\right) \tilde{a}^*\left(\bar{\bar{t}}{-}iT{-}\frac{\bar{m}T}{M}\right)\textrm{d}\bar{\bar{t}}}_{\substack{\textrm{T}^{+}_{2}}}, \notag 
\end{align}
and 
$\mathrm{nz}(\gamma)$ 
is zero if $\gamma \leq 0$, and one elsewhere.
We note that $\textrm{T}^0_{1}$ in $\Theta_{0}$ in \eqref{Equ:Orthonorm_D-ODDM_proof5} is $N$ when $\bar{n}=0$, and zero elsewhere. Similarly, $\textrm{T}^0_{2}$ in $\Theta_{0}$  is $1$ when $\bar{m}=0$, and zero elsewhere. Moreover, $\textrm{T}^{+}_{2}$ and $\textrm{T}^{-}_{2}$  in $\Theta^+_{i}$ and $\Theta^-_{i}$, respectively, are zero for all values of $i\geqslant 1$. Based on these, it can be concluded that for \textit{SC} with $m\geqslant 0$, $\Psi^{\textrm{D-ODDM}}(m,m',n,n')=\Theta_0=\delta(\bar{m})\delta(\bar{n})$.


For \textit{SC} with $m< 0$, following steps similar to those for \textit{SC} with $m\geqslant 0$, it can be shown that 
\begin{align}\label{Equ:Orthonorm_D-ODDM_proof7}
\Psi^{\textrm{D-ODDM}}(m,m',n,n')&=\Theta_0{+}\mathrm{nz}(K{-}1)\sum_{i=1}^{K{-}1} \Theta^+_{i}{+}\sum_{i=1}^{K} \Theta^-_{i}\notag\\
&=\Theta_0
=\delta(\bar{m})\delta(\bar{n}).
\end{align}
Based on \eqref{Equ:Orthonorm_D-ODDM_proof5} and \eqref{Equ:Orthonorm_D-ODDM_proof7}, it can be concluded that  for \textit{SC}, $\Psi^{\textrm{D-ODDM}}(m,m',n,n')=\delta(\bar{m})\delta(\bar{n})$.

We next focus on \textit{GC-I} where $T_{a1}=KT$ with $K\geq 0$ and $T_{a2}\leqslant \frac{T}{2}$. For $0 \leqslant m\leqslant 2Q$,  it can be shown that 
\begin{align}\label{Equ:Orthonorm_D-ODDM_proof8}
\Psi^{\textrm{D-ODDM}}(m,m',n,n')&=\Theta_0{+}\mathrm{nz}(K)\sum_{i=1}^K \Theta^+_{i}{+}\mathrm{nz}(K)\sum_{i=1}^{K} \Theta^-_{i}\notag\\
&=\Theta_0=\delta(\bar{m})\delta(\bar{n}).
\end{align} 
Similarly, for $2Q < m< M-2Q$ in \textit{GC-I}, 
\begin{align}\label{Equ:Orthonorm_D-ODDM_proof9}
&\Psi^{\textrm{D-ODDM}}(m,m',n,n')\notag\\
&=\Theta_0{+}\mathrm{nz}(K)\sum_{i=1}^K \Theta^+_{i}{+}\mathrm{nz}(K{-}1)\sum_{i=1}^{K{-}1} \Theta^-_{i}=0,
\end{align}
and for $M-2Q \leqslant m\leqslant M-1$ in \textit{GC-I}, 
\begin{align}\label{Equ:Orthonorm_D-ODDM_proof10}
&\Psi^{\textrm{D-ODDM}}(m,m',n,n')\notag\\
&=\Theta_0{+}\sum_{i=1}^{K+1} \Theta^+_{i}{+}\mathrm{nz}(K{-}1)\sum_{i=1}^{K{-}1} \Theta^-_{i}=0. 
\end{align}
Based on \eqref{Equ:Orthonorm_D-ODDM_proof8}-\eqref{Equ:Orthonorm_D-ODDM_proof10}, and similar expressions for $m<0$, it can be concluded that  for \textit{GC-I}, $\Psi^{\textrm{D-ODDM}}(m,m',n,n')=\delta(\bar{m})\delta(\bar{n})$.

We next focus on \textit{GC-II} where $T_{a1}=KT$ with $K\geq 0$ and $T_{a2}> \frac{T}{2}$. For $0 \leqslant m\leqslant M-2Q$, $\Psi^{\textrm{D-ODDM}}(m,m',n,n')$ is same as that in \eqref{Equ:Orthonorm_D-ODDM_proof8}, and for $2Q \leqslant m\leqslant M-1$, $\Psi^{\textrm{D-ODDM}}(m,m',n,n')$ is same as that in \eqref{Equ:Orthonorm_D-ODDM_proof10}. For $M-2Q< m< 2Q$,  it can be shown that  
\begin{align}\label{Equ:Orthonorm_D-ODDM_proof11}
&\Psi^{\textrm{D-ODDM}}(m,m',n,n')\notag\\
&=\Theta_0{+}\sum_{i=1}^{K+1} \Theta^+_{i}{+}\mathrm{nz}(K)\sum_{i=1}^{K} \Theta^-_{i}=0.
\end{align}
Based on these and similar expressions for $m<0$, it can be concluded that  for \textit{GC-II}, $\Psi^{\textrm{D-ODDM}}(m,m',n,n')=\delta(\bar{m})\delta(\bar{n})$. 
Finally, 
the results for \textit{GC-I} and \textit{GC-II} together guarantees~\eqref{Equ:Orthonorm_D-ODDM}, which completes the proof of Theorem \ref{Thr:Orthonorm_D-ODDM}.
\hfill $\blacksquare$ 


\end{document}